\documentclass[%
 reprint,
 amsmath,amssymb,
 aps,
]{revtex4-2}

\usepackage{graphicx}
\usepackage{dcolumn}
\usepackage{bm}

\begin{document}


\title{Field Analysis for a Highly-Overmoded Iris Line, \\ with Application to THz Radiation Transport at LCLS}

\author{Adham Naji}\email{anaji@stanford.edu}
 \altaffiliation[]{}
\author{Gennady Stupakov}
\author{Zhirong Huang}
\author{Karl Bane}
 
\affiliation{%
 SLAC National Linear Accelerator Laboratory, Stanford University, Menlo Park, CA 94025
}%

\date{\today}

\begin{abstract}
Vector field analysis is presented for a highly overmoded iris-line structure that can act as a medium of transportation for THz radiation. The axisymmetric structure is capable of supporting hybrid modes with desirable features such as low propagation loss,  uniformly linear polarization and approximately-Gaussian intensity profile across the iris. A specific application that can benefit from these desireable features is the transportation of THz undulator radiation over hundreds of meters to reach the experimental halls at the LCLS facility at SLAC, Stanford. Such a structure has been modelled before as a boundary-value problem using Vainstein's complex-impedance boundary condition and assuming infinitely-thin screens. Given that physical realizations of such screens must have finite thickness and that the THz wavelength in the 3--15~THz range is expected to be smaller than convenient screen thicknesses in practice, the question of the impact of finite screen thickness on propagation performance becomes rather pressing. To address this question, we present a mode-matching analysis of the structure as an open resonator with finite screen thickness and perturbatively clustered (localized) field expansions, for computational feasibility. The effect of screen thickness is seen to lower the attenuation constant on the iris line, which is dominated by diffraction loss. Ohmic loss due to the finite conductivity of metallic surfaces at the screen edges are found to be negligible compared to diffraction loss. The propagation loss predictions based on the Vainstein-model are compared with the numerical results from mode-matching for infinitely-thin screens, where the former method is observed to agree with the numerical results better at higher Fresnel numbers (highly-overmoded structures). The properties of the dominant mode fields are formally derived from first principles and a recommended approach is discussed for the inclusion of screen-thickness effects into propagation loss estimations. 
\end{abstract}

\maketitle



\section{\label{Intro}Introduction}

Efficient transportation of THz radiation over long distances (hundreds of meters) is a challenging problem.  On one hand, the THz wavelength (say, 20--100 $\mu$m, for the range 3--15 THz), being smaller than microwaves, makes the scaled transverse dimensions of a traditional single-moded waveguide impractically small over long distances. On the other hand,  being larger than optical wavelengths makes the THz wave relatively more prone to Fresnel diffraction.  An example of this challenge, which was the original motivator behind the present study, is the problem of efficiently transporting radiation from an ``afterburner" THz linear undulator downstream of LCLS over a distance of 150--350~m, to reach the experimental halls at the LCLS facility, SLAC, Stanford \cite{Zhang,DESY}. A traditional quasi-optical solution that utilizes a combination of planar, toroidal or paraboloidal mirrors to relay the THz beam in multiple steps is one proposed solution, which typically suffers from power loss of approximately 1\% per mirror as well as some aberration and misalignment \cite{Zhang,DESY2}. To reach the near experimental hall at LCLS, for example, through a 150-m path (roughly 34 mirrors) going through the access maze at LCLS, the mirrors are estimated to incur around 30\% power loss \cite{Zhang}. The iris-line structure, whose analysis is the subject of this paper, is an alternative solution that was first proposed by Geloni, \emph{et al.}, \cite{DESY} for the THz transport at LCLS. The iris line supports an attractive hybrid mode that can deliver (1) a low propagation loss, (2) an almost-Gaussian intensity profile and (3) an invariant linear polarization across the iris. These features make it ideal for direct coupling with the radiation incident from the THz linear undulator. 
\begin{figure}
\includegraphics[width=0.83\columnwidth]{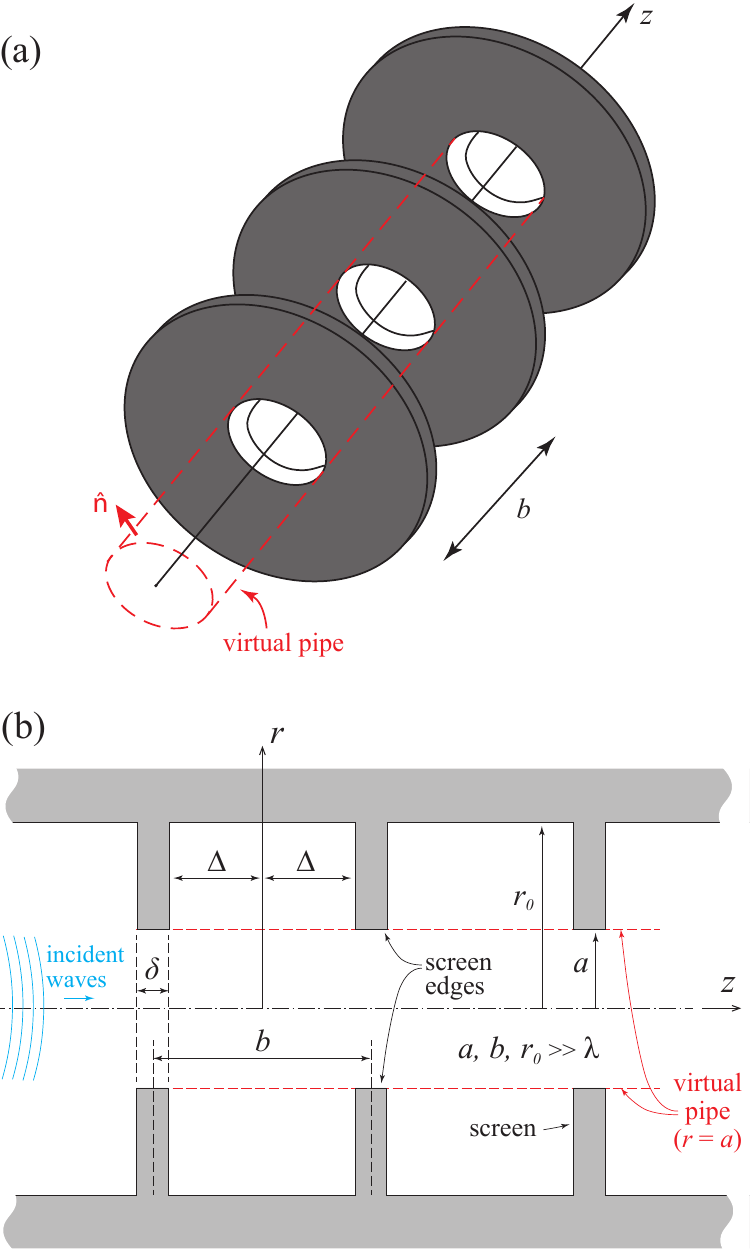}
\caption{\label{fig:geometry1}  (a) The iris-line geometry shown in three-dimensions, without the enclosing chamber at the outer radius. (b) A cross section in the iris line, showing dimensions and labeling.}
\end{figure}

The iris-line geometry under consideration consists of a series of parallel screens, each having a circular iris (gap) of radius $a$ and centered around the axis $z$, as shown in Figure~\ref{fig:geometry1}. The periodic structure has a period $b$ and iris radius $a$ that are much larger (hundreds or throusands) of the THz wavelength in vacuum, $\lambda_{0}$, making the structure highly overmoded. The screens expand transversely to an outer radius $r_{0}>a$ that defines the boundary of the enclosing chamber. The ``virtual pipe'' furnished by the irises along the axial ($z$) direction provides the resonant structure with a means to act as a waveguide (transmission line). On the other hand, the structure's resonance conditions, controlled by the periodicity and the shape of the annular regions sandwiched between the screens, will constitute equivalent boundary conditions from the perspective of the waveguide. 

\begin{figure}
\includegraphics[width=\columnwidth]{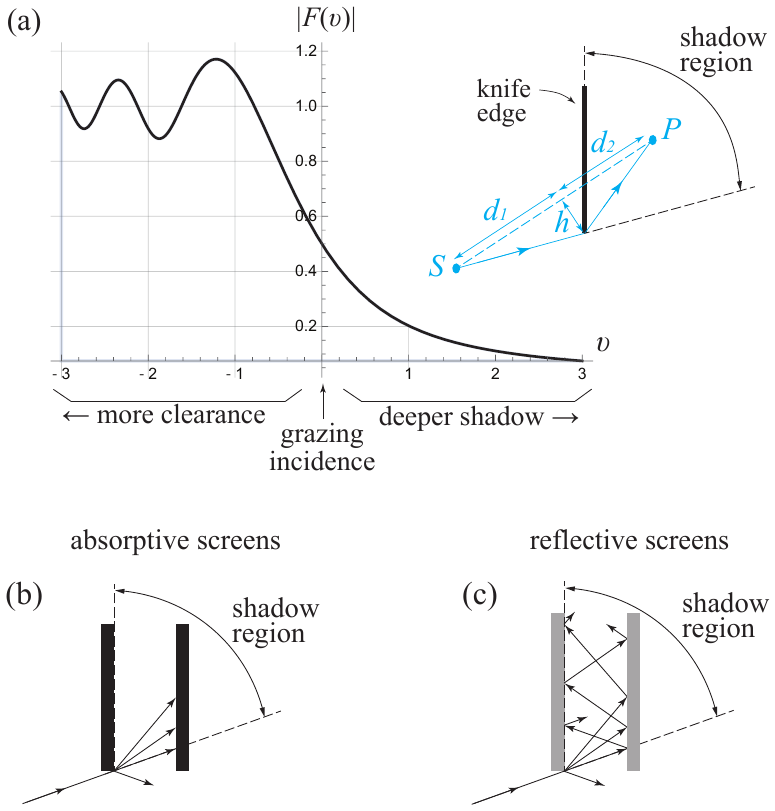}
\caption{\label{fig:geometry2} (a) A canonical example of Fresnel diffraction of a plane wave caused by a single knife-edge obstruction (thin and absorptive screen). The field strength of the wave traveling from point $S$ to point $P$ diminish according to the function $|F(\nu)|$ shown. As the point $P$ moves deeper into the shadow region behind the screen, the attenuation is increased. The Fresnel parameter $\nu=h\sqrt{2(d_{1}+d_{2})/(\lambda_{0}d_{1}d_{2})}$ defining the depth of shadow (or clearance) is found geometrically from the dimensions shown. (b) A simplified illustration of rays lost to diffraction, as they hit an absorptive screen or get diffracted into the shadow region bound by two such screens. (c) A simplified illustration of rays lost to reflection or diffraction in the region between two conductive screens. }
\end{figure}

For a plane wave that is paraxially incident on the iris line (Figure~\ref{fig:geometry1}), the Fresnel diffraction (or knife-edge diffraction) experienced at the screen edges will cause part of the wave to be lost into the shadow region between screens. For a single screen edge illuminated by a plane wave, this is similar to the well-known phenomenon of knife-edge obstruction (e.g.~\cite{Saunders}), which causes the diffracted field amplitude to decrease as we go deeper into the shadow  (see Figure~\ref{fig:geometry2}a) according to the attenuation factor $|F(\nu)|$, where $F(\nu)=\frac{1+i}{2}\int^{\infty}_{\nu}e^{-i\pi \tau^{2}/2}d\tau$ and $\nu$ is the Fresnel diffraction parameter defined as $\nu=h\sqrt{2(d_{1}+d_{2})/(\lambda_{0}d_{1}d_{2})}$.  Assuming that the screens are perfectly absorptive, all the diffracted waves entering the shadow will constitute lost power from the perspective of propagation along the axis. This can be equivalently modeled as an open-resonator structure, where we let the outer radius $r_{0}$ go to infinity and the screens can be assumed to be perfectly absorptive or conductive; in either case the same diffraction mechanism is seen at the screen edges and none of the diffracted waves entering the shadow region will return to the line, as shown Figures~\ref{fig:geometry2}b,c. Note that such an ideal open-resonator model (i.e.~with $r_{0}{\rightarrow}\infty$) can be well approximated in practice by making the distance $r_{0}{-}a$ finite but deep enough such that the power lost to diffraction is approximately equal to that lost in the limit $r_{0}{\rightarrow}\infty$. This is typically achieved by taking the depth $r_{0}{-}a$ to be much larger than the diffraction scale ${\sim}\sqrt{b \lambda_{0}}$, \cite{Chao}.

One way to analyze this open resonator is to consider the diffraction effects between the screens equivalent to a complex impedance boundary condition at $r=a$ and thereby convert the problem into a boundary-value problem over the closed virtual pipe of radius $r=a$, \cite{DESY}. Such a boundary condition has been called the Vainstein boundary condition \cite{DESY,Vainstein1,Vainstein2} and provides one of the methods of analysis we shall use to describe the losses on the iris line. This method, however, assumes infinitely-thin screens ($\delta=0$). In practice, the screens will have to be of finite thickness. Indeed, the finite thickness of the screen will be quite large relative to the wavelength at THz frequencies and a question is raised as to whether the screen thickness will influence the attractive features (low loss, invariant linear polarization and almost-Gaussian amplitude profile) of propagation using the iris line, as predicted by Vainstein's model. To address this question we analyze the open-resonator structure using the mode matching method and clustered (localized) field expansions, for computational feasibility. 

For this type of overmoded open-resonator structures, it will be seen that the dominant power loss mechanism is due to diffraction or radiation loss, not ohmic (conductive) loss. This should be compared with smooth waveguides and traditional corrugated structures, typically found at microwave frequencies where the depth and period of corrugations are shallow and often $\leq \lambda/4$ (e.g.~\cite{Mahmoud,StupakovCorrugated,Borgnis,Hutter,Clarricoats1,Clarricoats2}). For example, the dominant hybrid mode on the iris line will be shown to have the desirable property of the attenuation constant decreasing with increasing radius and increasing frequency, as a function of $a^{-3}\omega^{-3/2}$. The same property for the attenuation constant is found in smooth circular waveguides operating with the $\text{TE}_{0n}$ modes \cite{Mahmoud}. However, the attenuation on the iris line is mainly due to \emph{diffraction} loss, while on the smooth circular waveguide is due to \emph{ohmic} loss (by azimuthal surface currents) \cite{Mahmoud,Pozar}. 

Employing this iris line for THz radiation transport in practice will clearly require a study of not only its propagation power loss, polarization purity and amplitude profile, but also of critical considerations such as correct mode launching and coupling, input transients, dispersion and  tolerance to fabrication errors and mechanical misalignment. In this paper, we confine our scope to the investigation of propagation loss, polarization and amplitude profile properties, assuming screens of finite thickness and paraxial plane wave incidence (no electron beam on the line). 

The paper is organized as follows. In Section~\ref{Vainstein} we briefly review the Vainstein-based model and use its predictions as a reference for comparison. We then proceed to Section~\ref{mode-matching} where we derive the vector field equations for the dominant dipole mode of the iris line using the method of  mode matching and clustered expansions. In Section~\ref{results} we discuss the implementation of mode-matching and present its results for the specific example of the iris-line structure proposed for THz transport at LCLS, while investgiating the effect of finite screen thickness on the propagation properties. The paper concludes in Section~\ref{Conclusions}, followed by two Appendices that contain more formal derivations. Appendix~\ref{Appx1} presents the use of perturbation theory to formally derive the vector field properties of the hybrid mode on the iris line starting from Vainstein's model. Appendix~\ref{Appx2} uses perturbation theory to provide a justification for the use of the method of clustered field expansions in the analysis of overmoded paraxial iris lines.

Throughout this paper, we assume harmonic time dependence of the form $e^{-i\omega t}$.

\section{\label{Vainstein} Analysis based on Vainstein's complex boundary condition}
The open-resonator model of the iris line can be converted to an equivalent closed resonator problem which consists of Vainstein's approximate boundary condition at the cylindrical wall $r=a$. This boundary condition is formulated by Geloni, \emph{et al.}, in \cite{DESY} as
\begin{equation}
\left[ E + (1+i)\hat{\beta}_{0}aM \frac{\partial}{\partial r} E\right]_{r=a}={0}, \label{BCvectorial__}
\end{equation}
where $\hat{\beta}_{0}=0.824$, $M=1/\sqrt{8\pi N_{f}}$ with $N_{f}=a^{2}/(b\lambda_{0})$ being the Fresnel number, and $E$ represents the electric field envelope components $E_{r}$ or $E_{\theta}$ in the cylindrical frame of coordinates $(r,\theta,z)$. This analytical model assumes that the parameter $M$ is small ($M\ll 1$) and that the structure is highly-oversized compared to the wavelength, with $k_{0}b\gg 1$, where $k_{0}=2\pi/\lambda_{0}$ is the wavenumber in free-space. 
\hspace{0cm}	
\begin{figure}
\centering
\includegraphics[width=\columnwidth,angle=0]{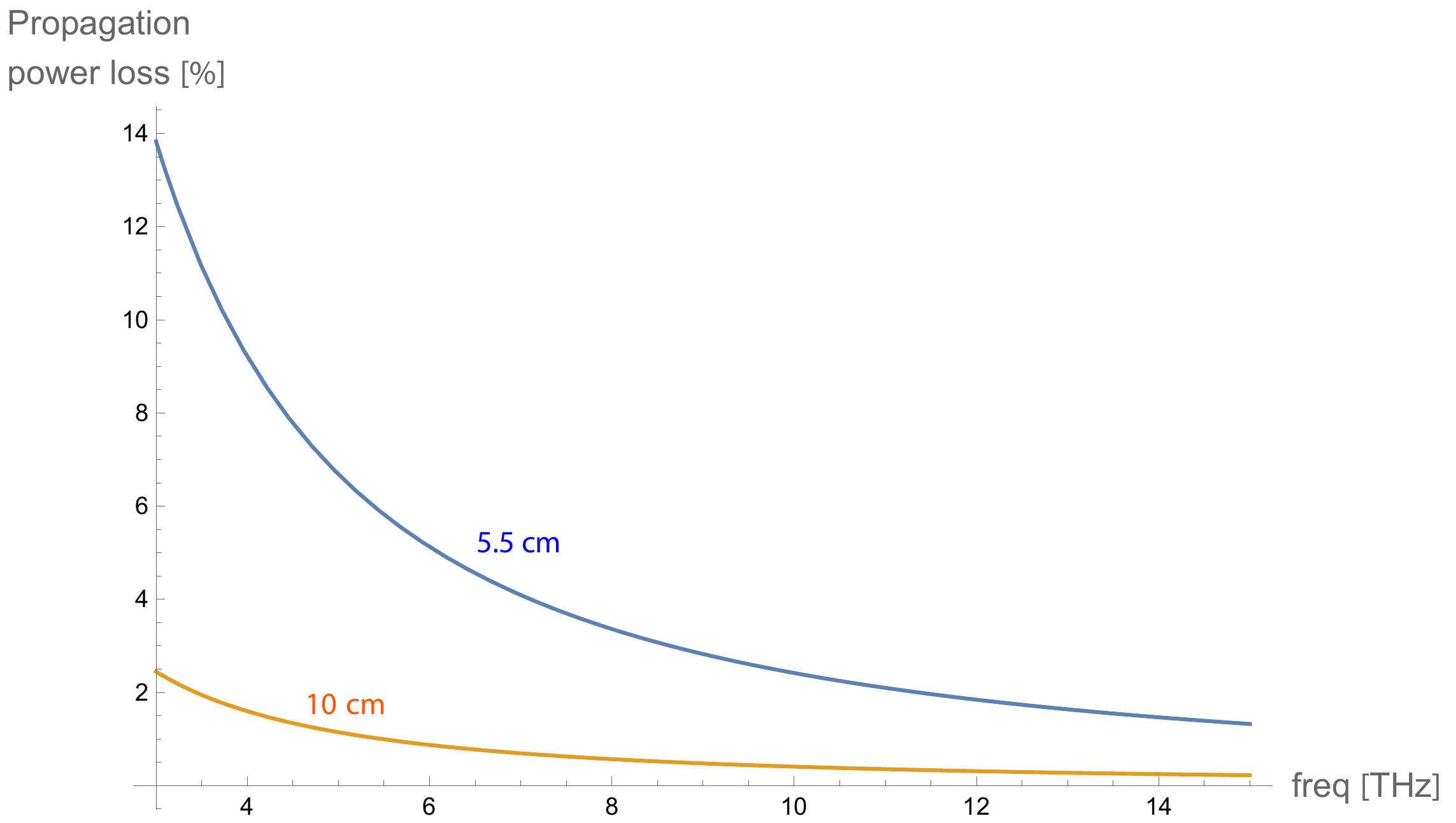}
\caption{Power loss for 150 m (to reach the Near Experimental Hall at LCLS) for the frequency range 3--15~THz, using an iris-line with a period of $b=30$ cm and two iris radii ($a=5.5$ or $10$ cm). This prediction is based on the Vainstein-model and assumes zero screen thickness. The larger radius is recommended for lower loss, if mechanically feasible.}   \label{fig:LossCurveFinalBefore}
\end{figure}

Solving the Helmholtz wave equation while imposing the boundary condition (\ref{BCvectorial__}) results in a dominant hybrid mode with invariant linear polarization across the iris, an amplitude profile approximately equal to the Bessel function $J_{0}(2.4r/a)$, and the following propagation power loss ($L_{p}$) law (see Appendix~\ref{Appx1} for derivations)
\begin{equation}
    L_{p}=\left[1-e^{-4.75 c^{3/2} b^{1/2} \omega^{-3/2}a^{-3}z} \right]\times 100\%, \label{VainsteinLoss__}
\end{equation}
where $c$ is the speed of light in vacuum, $\omega$ is the angular frequency and $z$ is the distance travelled down the line. A key feature of this loss law is the tendency of the attenuation constant $\text{Im}[\beta]$ to drop with larger radii and frequencies (inversely proportional to $a^{3}$ and $\omega^{3/2}$), where $\beta$ is the complex propagation constant along the $z$ direction.

An iris line with dimensions $a=5.5$~cm and $b=30$~cm, for example, has been proposed \cite{DESY} for the THz radiation transport at LCLS. Using the power loss estimation (\ref{VainsteinLoss__}), we see that a THz undulator radiation in the range 3--15~THz will experience maximum propagation loss of $14\%$ at 3 THz. The loss decreases at higher frequencies, as shown in Figure~\ref{fig:LossCurveFinalBefore}. If larger radii are feasible for installation at LCLS, lower loss can be achieved by exploiting the cubic-law dependence of loss on radius $a$. For example, Figure~\ref{fig:LossCurveFinalBefore} shows a second iris line with $a=10$ cm, and a power loss less than $3\%$ at 3~THz.

The predictions given by (\ref{VainsteinLoss__}) do not take into account the effect of finite screen thickness ($\delta>0$). In Section~\ref{mode-matching}, we analyse the modal fields on the line for nonzero screen thicknesses and compare the results with those predicted by Vainstein's model.

\section{\label{mode-matching} Analysis based on mode matching in an open resonator}

\begin{figure}
\includegraphics[width=0.8\columnwidth]{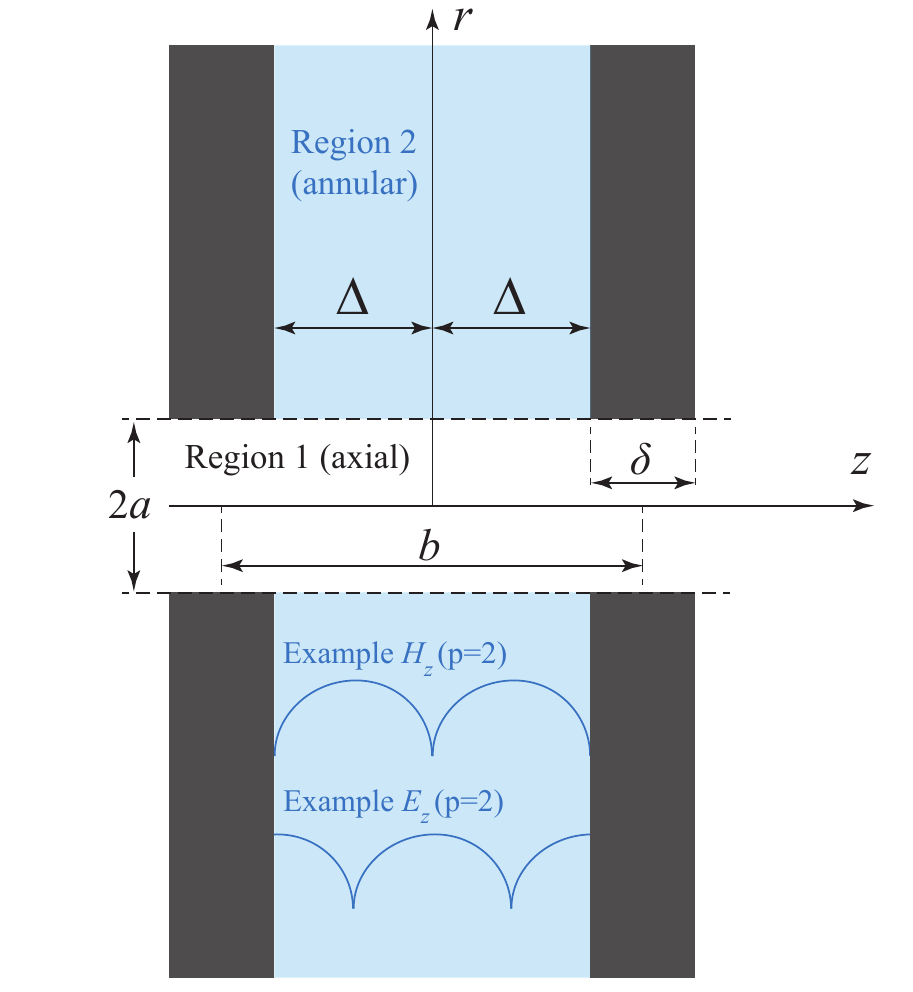}
\caption{\label{fig:geometry} Regions of analysis for the open-resonator model, shown over the cross section for one period (gap) in the iris-line structure.}
\end{figure}

In this analysis we begin by expanding the modal fields in the axial region (region 1 in Figure~\ref{fig:geometry}) and the annular region (region 2 in Figure~\ref{fig:geometry}), then impose the boundary conditions at $r=a$ and derive the structure's characteristic dispersion relation ($k_{0}$--$\beta$) in matrix form. Solving the characteristic equation for a given wavenumber $k_{0}=\omega/c$ will then yield the complex propagation constant $\beta$ on the iris-line. The presented approach is similar to the classical treatment given by Zotter and Bane \cite{Zotter} in terms of matrix equation conditioning. However, the present treatment is different from \cite{Zotter} in that it is concerned with the analysis of an open ($r_{0}{\rightarrow}\infty$), rather than closed, resonator model and in that we derive all the field components directly using the longitudinal components ($E_{z}, H_{z}$) of the fields themselves, rather than the Hertz potentials.  Recall from Whittaker's theorem \cite{Zangwill,jackson}, that only two independent solutions of the scalar wave equation are needed to determine all the 6 components of the EM field in vacuum. The two approaches are therefore fundamentally equivalent.  


\subsection{Field expansion in the axial region (region 1)} \label{region1}
Starting with Maxwell's curl equations, $\nabla\times\mathbf{E}=i\omega\mu\mathbf{H}$ and $\nabla\times\mathbf{H}=-i\omega\epsilon\mathbf{E}$, and seeking wave solutions that are travelling along $z$ in the form $e^{i\beta z}$, we can rewrite these curl equations with all the vectors and the operator $\nabla$ divided into their axial (along $z$) and transversal (subscripted with $t$) parts, as
\begin{eqnarray} 
(\mathbf{\nabla}_{t}+i\beta_{z}\hat{z})\times(\mathbf{E}_{t}+E_{z}\times\hat{z})&=&i\omega \mu (\mathbf{H}_{t}+H_{z}\hat{z})\label{eq:general},\\
(\mathbf{\nabla}_{t}+i\beta_{z}\hat{z})\times(\mathbf{H}_{t}+H_{z}\times\hat{z})&=&-i\omega \epsilon (\mathbf{E}_{t}+E_{z}\hat{z}). \label{eq:general2}
\end{eqnarray}

For a transverse electric (TE) mode, we substitute $E_{z}=0$ into (\ref{eq:general}) and (\ref{eq:general2}) and equate vector terms that are parallel to each other on each side of the equations. After algebraic vector manipulation, we obtain
\begin{eqnarray} 
\mathbf{H}_{t}&=&\frac{\beta}{\omega\mu}\hat{z}\times\mathbf{E}_{t}, \ \mathbf{\nabla}_{t}\times\mathbf{E}_{t}=i\omega\mu H_{z}\hat{z}, \ \mathbf{\nabla}_{t}\times\mathbf{H}_{t}=0,\label{eq:RegularTE}\\
\mathbf{H}_{t}&=&\frac{i\beta}{k^{2}_{t}}\mathbf{\nabla}_{t}H_{z}, \ \ \ \ \mathbf{E}_{t}=\frac{i\omega\mu}{k^{2}_{t}}\mathbf{\nabla}_{t}H_{z}\times\hat{z}.\label{eq:RegularTE2}
\end{eqnarray}

For a transverse magnetic (TM) mode ($H_{z}=0$), we similarly find
\begin{eqnarray} 
\mathbf{E}_{t}&=&-\frac{\beta}{\omega\epsilon}\hat{z}\times\mathbf{H}_{t}, \ \mathbf{\nabla}_{t}\times\mathbf{H}_{t}=-i\omega\epsilon E_{z}\hat{z}, \ \mathbf{\nabla}_{t}\times\mathbf{E}_{t}=0,\label{eq:RegularTM}\\
\mathbf{E}_{t}&=&\frac{i\beta}{k^{2}_{t}}\mathbf{\nabla}_{t}E_{z}, \ \ \ \ \mathbf{H}_{t}=\frac{-i\omega\epsilon}{k^{2}_{t}}\mathbf{\nabla}_{t}H_{z}\times\hat{z}.\label{eq:RegularTM2}
\end{eqnarray}

If the boundary conditions of a given structure provide a form of coupling (impedance) that links the TE and TM modes, a hybrid mode is generally produced as a combination of the TE and TM modes, with
\begin{eqnarray} 
\mathbf{E}_{t}&=&\frac{i \omega \mu}{k^{2}_{t}} \left[  \mathbf{\nabla}_{t}H_{z}\times \hat{z} + \frac{\beta}{\omega \mu}\mathbf{\nabla}_{t} E_{z}  \right],\label{fieldForGuide}\\
\mathbf{H}_{t}&=&\frac{i \omega \epsilon}{k^{2}_{t}} \left[ - \mathbf{\nabla}_{t}E_{z}\times \hat{z} + \frac{\beta}{\omega \epsilon}\mathbf{\nabla}_{t} H_{z}  \right].\label{fieldForGuide2}
\end{eqnarray}

Consider now our iris-line structure and the field description for region 1, which must generally meet the hybrid mode equations (\ref{fieldForGuide}) and (\ref{fieldForGuide2}), as well as the scalar transverse Helmholtz equations,  $\nabla^{2}_{t}E_{z}+k^{2}_{t}E_{z}=0$ and $\nabla^{2}_{t}H_{z}+k^{2}_{t}H_{z}=0$, where $k_{t}=\sqrt{k_{0}^{2}-\beta^{2}}$ is the transverse wavenumber. Since both the Helmholtz equation and the present boundary conditions are separable in cylindrical coordinates \cite{Morse}, we can immediately assume that $E_{z}$ and $H_{z}$ are of the form $E_{z}=C R(r)\Theta(\theta)Z(z)$ and $H_{z}=D\hat{R}(r)\hat{\Theta}(\theta)\hat{Z}(z)$, where $C$ and $D$ are arbitrary constants. We find the form of the functions $R(r), \Theta(\theta)$ and $Z(z)$ that constitute $E_{z}$ as follows (with a similar treatment for $H_{z}$). Given the periodicity of the structure, $Z(z)$ is predicted by Floquet's theorem to have the form of a Bloch wave \cite{Collin,slaterbook}, whose field will be periodic from one structural period to the next, except for a complex phase advance of $e^{i\beta_{0}b}$, where $\beta_{0}$ is the propagation constant along $z$. Note that $\beta_{0}$ is generally complex to allow for attenuation in lossy structures (as in our iris-line structure), but would be real when the structure exhibit no losses to materials or radiation (i.e.~closed structure with perfectly conducting walls and perfect dielectrics) \cite{slaterbook}.  We can therefore write $Z(z)$ as $Z(z)=e^{i\beta_{0}z}\phi_{b}(z)$, where $\phi_{b}(z)$ is a periodic function of fundamental period $b$. Expanding $\phi(z)$ as a spacial Fourier series, $\sum_{n} C_{n}e^{i2n\pi z/b}$, leads to
\begin{equation} \label{Floquet}
Z(z)=\sum\limits^{\infty}_{n=-\infty}C_{n}e^{i\beta_{n}z}, \ \ \ \beta_{n}=\beta_{0}+2\pi n/b,
\end{equation}
where $\beta_{n}$ is the propagation constant and $k_{tn}=\sqrt{k^{2}_{0}-\beta_{n}^{2}}$ is the transverse wavenumber for the $n^{\text{th}}$ spacial harmonic. As in (\ref{fieldForGuide}) and (\ref{fieldForGuide2}), the transverse components of the $n^{\text{th}}$ harmonic can be found from its $E_{zn}, H_{zn}$ components as
\begin{eqnarray}
\mathbf{E}_{tn}&=&\frac{i \omega \mu}{k^{2}_{tn}} \left[  \mathbf{\nabla}_{t}H_{zn}\times \hat{z} + \frac{\beta_{n}}{\omega \mu}\mathbf{\nabla}_{t} E_{zn}  \right], \label{fieldForGuide_n}\\
\mathbf{H}_{tn}&=&\frac{i \omega \epsilon}{k^{2}_{tn}} \left[ - \mathbf{\nabla}_{t} E_{zn}\times \hat{z} + \frac{\beta_{n}}{\omega \epsilon}\mathbf{\nabla}_{t} H_{zn}  \right]. \label{fieldForGuide_n2}
\end{eqnarray}

To find the radial and azimuthal dependences, $R$ and $\Theta$, we now write the Helmholtz equation $\nabla^{2}_{t}E_{zn}+k^{2}_{tn}E_{zn}=0$ in cylindrical coordinates. Since the transverse Laplacian operator [$\nabla^{2}_{t}\equiv\frac{1}{r}\frac{\partial}{\partial r}\left( r \frac{\partial}{\partial r} \right)+\frac{1}{r^{2}}\frac{\partial^{2}}{\partial \theta^{2}}$] in such coordinates couples the radial and azimuthal components without coupling the axial component, the $Z(z)$ function drops out of the equation and we end up with the following equation, after separating the variables,
\begin{equation} \label{eq:separation0}
r^{2}\frac{R''}{R}+r\frac{R'}{R}+r^{2}k^{2}_{tn}=-\frac{\Theta''}{\Theta}=m^{2},
\end{equation}
where $m$ is a constant integer (to keep the azimuthal dependence a single-valued function of $\theta$). This gives azimuthal dependence $\Theta$ in the form of $\cos m\theta$ or $\sin m\theta$. With no loss of generality, we use the former for $E_{zn}$ and the latter for $H_{zn}$, since the azimuthal dependence $\hat{\Theta}$ in $H_{z}$ will turn out (as it must) to be of the same form. We also anticipate that $m$ will be the same across region 1 and 2, to maintain phase matching and as it does not depend on $n$; we hence reuse the same symbol $m$ in both regions. The radial dependence is now seen to reduce to
\begin{equation} \label{eq:Bessel0}
r^{2}R''+rR'+[r^{2}k^{2}_{tn}-m^{2}]R=0,
\end{equation}
which is the parameterized Bessel equation. Since the radius in region 1 is bounded by the iris radius ($r\leq a$) and includes the axis ($r=0$), we choose our solution to be the Bessel function of the first kind, $J_{m}(k_{tn}r)$, and reject the second kind. With a similar treatment for $H_{zn}$, we now have 
\begin{eqnarray} 
E_{zn}&=&\cos m\theta \sum\limits^{\infty}_{n=-\infty} C_{n} J_{m}(k_{tn}r)e^{i\beta_{n}z},\label{eq:EzHzLine}\\
H_{zn}&=&\sin m\theta \sum\limits^{\infty}_{n=-\infty} \frac{D_{n}}{Z_{0}} J_{m}(k_{tn}r)e^{i\beta_{n}z},\label{eq:EzHzLine2}
\end{eqnarray}
where, for symmetry, we have chosen to give $C$ and $D$ the same physical dimensions and explicitly isolate the free-space impedance, $Z_{0}=\sqrt{\mu/\epsilon}$.  We now substitute (\ref{eq:EzHzLine}) and (\ref{eq:EzHzLine2}) into (\ref{fieldForGuide_n}) and (\ref{fieldForGuide_n2}) to yield, after algebraic manipulation, the hybrid field expressions for region 1, as
\begin{widetext}
\begin{eqnarray} \label{TotalHybridLineFINAL}
E_{z_\text{I}}&=&\cos m\theta \sum\limits_{n=-\infty}^{\infty}C_{n} J_{m}(k_{tn}r)e^{i\beta_{n}z},\\
H_{z_\text{I}}&=&\sin m\theta \sum\limits_{n=-\infty}^{\infty}\frac{D_{n}}{Z_{0}} J_{m}(k_{tn}r)e^{i\beta_{n}z},\\
E_{r_\text{I}}&=&i\cos m\theta \sum\limits_{n=-\infty}^{\infty} \left[  \frac{D_{n}}{Z_{0}} \frac{\omega\mu m}{rk^{2}_{tn}}J_{m}(k_{tn}r) + C_{n} \frac{\beta_{n}}{k_{tn}} J'_{m}(k_{tn}r) \right]e^{i\beta_{n}z},\\
E_{\theta_\text{I}}&=&i\sin m\theta \sum\limits_{n=-\infty}^{\infty} \left[ - \frac{D_{n}}{Z_{0}} \frac{\omega\mu}{k_{tn}}J'_{m}(k_{tn}r) - C_{n} \frac{\beta_{n} m}{r k^{2}_{tn}} J_{m}(k_{tn}r) \right]e^{i\beta_{n}z},\\
H_{r_\text{I}}&=&i\sin m\theta \sum\limits_{n=-\infty}^{\infty} \left[  \frac{D_{n}}{Z_{0}} \frac{\beta_{n}}{k_{tn}}J'_{m}(k_{tn}r) + C_{n} \frac{\omega\epsilon m}{r k^{2}_{tn}} J_{m}(k_{tn}r) \right]e^{i\beta_{n}z},\\
H_{\theta_\text{I}}&=&i\cos m\theta \sum\limits_{n=-\infty}^{\infty} \left[  \frac{D_{n}}{Z_{0}} \frac{\beta_{n}m}{rk^{2}_{tn}}J_{m}(k_{tn}r) + C_{n} \frac{\omega\epsilon}{k_{tn}} J'_{m}(k_{tn}r) \right]e^{i\beta_{n}z},
\end{eqnarray}
\end{widetext}
where $J_{m}'[\cdot]$ denotes the derivative of $J_{m}[\cdot]$ with respect to its argument. Note that these equations will be further normalized following mode-matching in Section~\ref{modeMatch}, to provide symmetry and convenience during calculation.


\subsection{Field expansion in the annular region (region 2)}
Since the field propagation in region 1 will vary periodically along $z$ with the same fundamental period as the structure's, we know that the field in one annular gap will be the same as that in all gaps. Therefore, we merely need to solve for one gap. Since the gap is bounded in the $z$ direction and open in the $r$ direction, we expect it to host a standing wave along the longitudinal direction and a traveling wave in the radial direction (radiating outwards). The latter is expected to meet the usual radiation boundary condition ($E,H$ $\rightarrow 0$) at infinity. For the fields $E_{z},H_{z}$ in the gap, we also expect to superimpose solutions of the form $e^{\pm i\beta_{p} z}$ in order to form the standing waves. These fields must satisfy the transverse Helmholtz wave equations,  $\nabla_{t}E_{z}+k^{2}_{tp}E_{z}=0$ and $\nabla_{t}H_{z}+k^{2}_{tp}H_{z}=0$, where $k_{tp}=\sqrt{k_{0}^{2}-\beta_{p}^{2}}$ and the index $p$ denotes a given mode (standing wave) in region 2. Since both the Helmholtz equation and the structure's boundary conditions in region 2 are separable in the cylindrical frame \cite{Morse}, we follow a treatment similar to the one used above for region 1 and we assume that $E_{z}=A_{p}R(r)\Theta(\theta)Z(z)$ and $H_{z}=B_{p}\hat{R}(r)\hat{\Theta}(\theta)\hat{Z}(z)$, where $A, B$ are arbitrary constants. 

For the $p^{\text{th}}$ TE mode, considering that $H_{z}$ must have nodes at the screen walls (see Figure~\ref{fig:geometry}), then $\hat{Z}_{p}(z)$ must be of the form $\hat{Z}_{p}(z)=\sin \beta_{p}(z+\Delta)=\frac{1}{2i}[e^{i\beta_{p}(z+\Delta)}-e^{-i\beta_{p}(z+\Delta)}]$, where $\beta_{p}=p\pi/(2\Delta)$. Similarly, for the $p^{\text{th}}$ TM mode, considering that $E_{z}$ must have maxima at the screen walls, $Z_{p}(z)$ must be of the form $Z_{p}(z)=\cos \beta_{p}(z+\Delta)=\frac{1}{2}[e^{i\beta_{p}(z+\Delta)}+e^{-i\beta_{p}(z+\Delta)}]$. We now note that equations (\ref{fieldForGuide}) and (\ref{fieldForGuide2}), which related the transverse components to the longitudinal fields ($E_{z}, H_{z}$), were originally derived for a travelling wave (waveguide setup) rather than a standing wave (gap setup), and will need to be adjusted accordingly. For the TE ($E_{z}=0$) case, we now need to take $\partial/\partial z\equiv i\beta_{p}$ for the $e^{i\beta_{p}(z+\Delta)}$ term and $\partial/\partial z\equiv -i\beta_{p}$ for the $e^{-i\beta_{p}(z+\Delta)}$ term; and similarly for the TM mode. Repeating the derivation steps taken for region 1 by separating Maxwell's curl equations in terms of the longitudinal and transverse parts for the $p^{\text{th}}$ mode, followed by algebraic vector manipulation, gives the following relations for the TE mode
\begin{eqnarray} 
H_{zp}&=&B_{p}\sin\beta_{p}(z+\Delta)\hat{R}_{p}(r)\hat{\Theta}_{p}(\theta), \label{eq:TEinSlot}\\
\mathbf{H}_{tp}&=&B_{p}\frac{\beta_{p}}{k^{2}_{tp}}\cos\beta_{p}(z+\Delta)\mathbf{\nabla}_{t}\left[\hat{R}_{p}(r)\hat{\Theta}_{p}(\theta)\right], \label{eq:TEinSlot2}\\
\mathbf{E}_{tp}&=&B_{p}\frac{i\omega\mu}{k^{2}_{tp}}\sin\beta_{p}(z+\Delta)\mathbf{\nabla}_{t}\left[\hat{R}_{p}(r)\hat{\Theta}_{p}(\theta)\right]\times\hat{z}.\label{eq:TEinSlot3}
\end{eqnarray}

In a similar manner, we have for the TM mode
\begin{eqnarray} 
E_{zp}&=&A_{p}\cos\beta_{p}(z+\Delta)R_{p}(r)\Theta_{p}(\theta), \label{eq:TMinSlot}\\
\mathbf{E}_{tp}&=&-A_{p}\frac{\beta_{p}}{k^{2}_{tp}}\sin\beta_{p}(z+\Delta)\mathbf{\nabla}_{t}\left[R_{p}(r)\Theta_{p}(\theta)\right], \label{eq:TMinSlot2}\\
\mathbf{H}_{tp}&=&-A_{p}\frac{i\omega\epsilon}{k^{2}_{tp}}\cos\beta_{p}(z+\Delta)\mathbf{\nabla}_{t}\left[R_{p}(r)\Theta_{p}(\theta)\right]\times\hat{z}.\label{eq:TMinSlot3}
\end{eqnarray}

To find the radial and  azimuthal functions ($R,\Theta$) in this region, we substitute into the Helmholtz equation and use the cylindrical Laplacian, as was done in region 1. This leads to the same azimuthal dependence (namely, $\cos m\theta$ or $\sin m\theta$) and the same parameterized Bessel equation (\ref{eq:Bessel0}). For its Bessel solutions in region 2, we pick the Hankel function of the first kind,  $H^{(1)}_{m}(k_{tp}r)$, since here we have an outwardly travelling wave and since, for a time dependence $e^{-i\omega t}$, the second kind Hankel would represent inward travel (with a similar treatment for the TM mode). The nature of the travelling wave implied can be easily revealed by writing down the definition of $H_{m}^{(1)}$ at large distances \cite{NIST}, viz.
\begin{equation} \label{eq:HankelLimit}
R=H_{m}^{(1)}(k_{tp}r)\sim \sqrt{\frac{2}{\pi k_{tp}r}} e^{ik_{tp}r} e^{i\pi m/2}e^{i\pi/4}.
\end{equation}

Substituting the $R,\Theta,\hat{R},\hat{\Theta}$ forms back into equations (\ref{eq:TEinSlot})--(\ref{eq:TMinSlot3}), we can now explicitly write the full hybrid field expressions for region 2 as 
\begin{widetext}
\begin{eqnarray} 
E_{z_\text{II}}&=&\cos m\theta \sum\limits_{p=0}^{\infty} A_{p}\cos\beta_{p}(z+\Delta)H_{m}^{(1)}(k_{tp}r), \label{eq:TotalHybridSlotFINAL}\\
H_{z_\text{II}}&=&\sin m\theta \sum\limits_{p=0}^{\infty}\frac{B_{p}}{Z_{0}}\sin\beta_{p}(z+\Delta)H_{m}^{(1)}(k_{tp}r), \label{eq:TotalHybridSlotFINAL2}\\
\mathbf{E}_{r_\text{II}}&=&\cos m\theta\sum\limits_{p=0}^{\infty}\frac{\sin\beta_{p}(z+\Delta)}{k^{2}_{tp}}\left[\frac{B_{p}}{Z_{0}} \frac{i\omega\mu m}{r}H_{m}^{(1)}(k_{tp}r) - A_{p} \beta_{p} k_{tp}H_{m}^{'(1)}(k_{tp}r)  \right], \label{eq:TotalHybridSlotFINAL3}\\
\mathbf{E}_{\theta_\text{II}}&=&\sin m\theta\sum\limits_{p=0}^{\infty}\frac{\sin\beta_{p}(z+\Delta)}{k^{2}_{tp}}\left[ -\frac{B_{p}}{Z_{0}} i\omega\mu k_{tp} H_{m}^{'(1)}(k_{tp}r) + \frac{A_{p}\beta_{p} m}{r} H_{m}^{(1)}(k_{tp}r)  \right], \label{eq:TotalHybridSlotFINAL4}\\
\mathbf{H}_{r_\text{II}}&=&\sin m\theta\sum\limits_{p=0}^{\infty}\frac{\cos\beta_{p}(z+\Delta)}{k^{2}_{tp}}\left[ \frac{B_{p}}{Z_{0}} \beta_{p} k_{tp} H_{m}^{'(1)}(k_{tp}r) + \frac{A_{p}i\omega\epsilon m}{r} H_{m}^{(1)}(k_{tp}r)  \right], \label{eq:TotalHybridSlotFINAL5}\\
\mathbf{H}_{\theta_\text{II}}&=&\cos m\theta\sum\limits_{p=0}^{\infty}\frac{\cos\beta_{p}(z+\Delta)}{k^{2}_{tp}}\left[  \frac{B_{p}}{Z_{0}}\frac{\beta_{p} m}{r}H_{m}^{(1)}(k_{tp}r) + A_{p} i\omega\epsilon k_{tp} H_{m}^{'(1)}(k_{tp}r)  \right], \label{eq:TotalHybridSlotFINAL6}
\end{eqnarray}
\end{widetext}
where $B_{0}=0$. Note that, for symmetry in equations and parameters, we take the impedance $Z_{0}$ explicitly out of the $B_{p}$ coefficients. These equations will be further normalized following mode-matching in Section~\ref{modeMatch}, to provide symmetry and convenience during calculation.

It is important to bear in mind that the fields (\ref{eq:TotalHybridSlotFINAL})--(\ref{eq:TotalHybridSlotFINAL6}) are givens for $|z|\leq\Delta$ inside the gap and are identically $0$ in the region $\Delta<|z|\leq\Delta+\delta/2$, where the screens are assumed to be made of a perfect conductor (see Figure~\ref{fig:geometry}). 

\subsection{Mode matching at $r=a$} \label{modeMatch}
We now solve for the coefficient families $(A_{p},B_{p})$ and $(C_{n},D_{n})$ by enforcing the continuity of the tangential $E$ and $H$ fields across the boundary $r=a$ (the continuity of the normal fields follows automatically from Maxwell's divergence equations). Namely, we require the four conditions at $r=a$
\begin{eqnarray}
(1) && \ E_{\theta_\text{I}}=E_{\theta_\text{II}}, \ \ \ (2) \  E_{z_\text{I}}=E_{z_\text{II}},  \nonumber\\ 
(3) && \  H_{\theta_\text{I}}=H_{\theta_\text{II}}, \ \ \ (4) \  H_{z_\text{I}}=H_{z_\text{II}}.  \label{BCs}
\end{eqnarray}

Enforcing the first condition gives us
\begin{widetext}
\begin{eqnarray} 
\sum\limits_{n} && \left[ \frac{D_{n}}{Z_{0}} \frac{-i\omega\mu}{k_{tn}} J'_{m}(k_{tn}a)-C_{n}\frac{i\beta_{n}m}{a k_{tn}^{2}} J_{m}(k_{tn}a) \right] e^{i\beta_{n}z} \nonumber \\
 &=& \begin{cases}
    \sum\limits_{p}\left[ -\frac{B_{p}}{Z_{0}}i\omega\mu k_{tp}H^{'(1)}_{m}(k_{tp}a) + A_{p} \frac{\beta_{p} m}{a} H^{(1)}_{m}(k_{tp}a) \right]\frac{\sin \beta_{p}(z+\Delta)}{k^{2}_{tp}},& \text{for } |z|\leq \Delta \\
    0,              & \text{for } \Delta<|z|\leq\Delta+\delta/2
\end{cases}
\end{eqnarray}
\end{widetext}

Multiplying both sides by $e^{-i\beta_{n}z}$ and integrating through $\frac{1}{b}\int^{b/2}_{-b/2}$, using integration by parts twice, this yields (after manipulation and abbreviating the obvious Bessel argument) equations (\ref{eq:tempp1}) and (\ref{eq:tempp2}) below. Following similar steps, we enforce the second, third and fourth conditions, using integration by parts twice for each case, to also get equations (\ref{eq:tempp3})--(\ref{eq:tempp5}).

\begin{widetext}
\begin{eqnarray} 
&& \frac{1}{b}\int\limits^{b/2}_{-b/2}e^{-i\beta_{n}z}\sin \beta_{p}(\Delta+z)dz =\frac{-i2\beta_{p}}{\Delta(\beta_{n}^{2}-\beta_{p}^{2})} \begin{cases}
\sin\beta_{n}\Delta, & \text{for even }p, \\
-i\cos\beta_{n}\Delta, & \text{for odd }p, 
\end{cases} \label{eq:tempp1}\\
\Rightarrow \frac{D_{n}}{Z_{0}}\frac{-i\omega \mu J'_{m}}{k_{tn}}&+&C_{n}\frac{-i\beta_{n} m J{m}}{ak^{2}_{tn}} =\nonumber\\
&& -i\sum\limits_{p}\left( \frac{B_{p}}{Z_{0}} \frac{-i\omega\mu H_{m}^{'(1)}}{ k_{tp}}+A_{p} \frac{\beta_{p}mH{m}^{(1)}}{ak^{2}_{tp}} \right)\underbrace{\frac{2\beta_{p}}{b(\beta_{n}^{2}-\beta^{2}_{p})}\begin{cases}
\sin\beta_{n}\Delta, & \text{for even }p, \\
-i\cos\beta_{n}\Delta, & \text{for odd }p, \label{eq:tempp2}
\end{cases}}_{M_{np}}
\end{eqnarray}
\begin{equation} 
C_{n}=\sum\limits_{p}A_{p} \frac{H_{m}^{(1)}}{J_{m}} \underbrace{\frac{2\beta_{n}}{b(\beta_{n}^{2}-\beta_{p}^{2})} \begin{cases}
\sin\beta_{n}\Delta, & \text{for even }p, \\
-i\cos\beta_{n}\Delta, & \text{for odd }p, 
\end{cases}}_{N_{np}} \label{eq:tempp3}
\end{equation}
\begin{equation} 
B_{p}=\sum\limits_{n}\frac{i D_{n}}{Z_{0}} \frac{J_{m}}{H_{m}^{(1)}} \underbrace{\frac{2\beta_{p}}{\Delta(\beta_{n}^{2}-\beta_{p}^{2})} \begin{cases}
\sin\beta_{n}\Delta, & \text{for even }p, \\
i\cos\beta_{n}\Delta, & \text{for odd }p, 
\end{cases}}_{\frac{b}{\Delta}M^{\dagger}_{pn}} \label{eq:tempp4}
\end{equation}
\begin{eqnarray} 
\frac{1+\delta_{p0}}{k^{2}_{tp}}&&\left(\frac{B_{p}}{Z_{0}}\frac{\beta_{p} m H_{m}^{(1)}}{a}+A_{p} i\omega\epsilon k_{tp}  H{m}^{'(1)}\right) \nonumber\\
&=& \sum\limits_{n}\left( C_{n} \frac{\omega\epsilon J'_{m}}{k_{tn}}+\frac{D_{n}}{Z_{0}} \frac{\beta_{n}m J{m}}{ak^{2}_{tn}} \right)\underbrace{\frac{2\beta_{n}}{\Delta(\beta_{n}^{2}-\beta^{2}_{p})}\begin{cases}
\sin\beta_{n}\Delta, & \text{for even }p, \\
i\cos\beta_{n}\Delta, & \text{for odd }p, 
\end{cases}}_{\frac{b}{\Delta}N^{\dagger}_{pn}}\label{eq:tempp5}
\end{eqnarray}
\end{widetext}
where the dagger symbol represents Hermitian conjugation and $\delta_{p0}$ is Kronecker's delta (equal to 1 when $p=0$ and zero otherwise).  Note that $M$ and $N$ will denote matrices that contain elements such as $M_{np}$ or $N_{np}$.

Further simplification and symmetry in the equations are achieved if we momentarily define $B'_{p}\equiv iB_{p}$ and remove the imaginary $i$ factor from odd cases in the $M, N$ matrices by absorbing it into the corresponding $\bar{A}_{p}, \bar{B}_{p}$ coefficients, calling the new matrixed $\bar{M}, \bar{N}$, as follows
\begin{eqnarray} 
\bar{A}_{p}&=&\begin{cases}
A_{p}, & \text{for even }p \\
-iA_{p}, & \text{for odd }p \\
\end{cases} \\
\bar{B}_{p}&=&\begin{cases}
B'_{p}, & \text{for even }p \\
-iB'_{p}, & \text{for odd }p \\
\end{cases}=\begin{cases}
i B_{p}, & \text{for even }p \\
B_{p}, & \text{for odd }p \\
\end{cases}\\
\bar{M}_{np}&=&\frac{2\beta_{n}}{b(\beta_{n}^{2}-\beta_{p}^{2})}\begin{cases}
\sin\beta_{n}\Delta, & \text{for even }p \\
\cos\beta_{n}\Delta, & \text{for odd }p \\
\end{cases} \\
\bar{N}_{np}&=&\frac{2\beta_{p}}{b(\beta_{n}^{2}-\beta_{p}^{2})}\begin{cases}
\sin\beta_{n}\Delta, & \text{for even }p \\
\cos\beta_{n}\Delta, & \text{for odd }p \\
\end{cases} 
\end{eqnarray}

This would render the equations real, since the coefficients themselves are arbitrary. For symmetry and convenience, we also normalize the coefficients as $J_{m}(k_{tn}a)C_{n}\rightarrow C_{n}$, $J_{m}(k_{tn}a)D_{n}\rightarrow D_{n}$, $H^{(1)}_{m}(k_{tp}a)A_{p}\rightarrow A_{p}$ and $H^{(1)}_{m}(k_{tp}a)B_{p}\rightarrow B_{p}$, which is equivalent to dividing the original mode equations in  (\ref{TotalHybridLineFINAL}) and (\ref{eq:TotalHybridSlotFINAL}) by $J_{m}(k_{tn}a)$ for all the Bessel terms and their derivatives, whilst dividing by $H^{(1)}_{m}$ for all the Hankel terms and their derivatives. Effecting this normalization gives us the final field equations in region 1 and 2 as follows (noting that $B_{0}=0$)
\begin{widetext}
\begin{eqnarray} 
E_{z_\text{I}}&=&\cos m\theta \sum\limits_{n=-\infty}^{\infty}C_{n} \frac{J_{m}(k_{tn}r)}{J_{m}(k_{tn}a)}e^{i\beta_{n}z}\label{TotalHybridLineFINALnormalized}\\
H_{z_\text{I}}&=&\sin m\theta \sum\limits_{n=-\infty}^{\infty}\frac{D_{n}}{Z_{0}} \frac{J_{m}(k_{tn}r)}{J_{m}(k_{tn}a)}e^{i\beta_{n}z}\label{TotalHybridLineFINALnormalized2}\\
E_{r_\text{I}}&=&i\cos m\theta \sum\limits_{n=-\infty}^{\infty} \left[  \frac{D_{n}}{Z_{0}} \frac{\omega\mu m}{rk^{2}_{tn}}\frac{J_{m}(k_{tn}r)}{J_{m}(k_{tn}a)} + C_{n} \frac{\beta_{n}}{k_{tn}} \frac{J'_{m}(k_{tn}r)}{J_{m}(k_{tn}a)} \right]e^{i\beta_{n}z}\label{TotalHybridLineFINALnormalized3}\\
E_{\theta_\text{I}}&=&i\sin m\theta \sum\limits_{n=-\infty}^{\infty} \left[ - \frac{D_{n}}{Z_{0}} \frac{\omega\mu}{k_{tn}}\frac{J'_{m}(k_{tn}r)}{J_{m}(k_{tn}a)} - C_{n} \frac{\beta_{n} m}{r k^{2}_{tn}} \frac{J_{m}(k_{tn}r)}{J_{m}(k_{tn}a)} \right]e^{i\beta_{n}z}\label{TotalHybridLineFINALnormalized4}\\
H_{r_\text{I}}&=&i\sin m\theta \sum\limits_{n=-\infty}^{\infty} \left[  \frac{D_{n}}{Z_{0}} \frac{\beta_{n}}{k_{tn}}\frac{J'_{m}(k_{tn}r)}{J_{m}(k_{tn}a)} + C_{n} \frac{\omega\epsilon m}{r k^{2}_{tn}} \frac{J_{m}(k_{tn}r)}{J_{m}(k_{tn}a)} \right]e^{i\beta_{n}z}\label{TotalHybridLineFINALnormalized5}\\
H_{\theta_\text{I}}&=&i\cos m\theta \sum\limits_{n=-\infty}^{\infty} \left[  \frac{D_{n}}{Z_{0}} \frac{\beta_{n}m}{rk^{2}_{tn}}\frac{J_{m}(k_{tn}r)}{J_{m}(k_{tn}a)} + C_{n} \frac{\omega\epsilon}{k_{tn}} \frac{J'_{m}(k_{tn}r)}{J_{m}(k_{tn}a)} \right]e^{i\beta_{n}z}\label{TotalHybridLineFINALnormalized6}
\end{eqnarray}
\begin{eqnarray} 
E_{z_\text{II}}&=&\cos m\theta\sum\limits_{p=0}^{\infty} A_{p}\cos\beta_{p}(z+\Delta)\frac{H_{m}^{(1)}(k_{tp}r)}{H_{m}^{(1)}(k_{tp}a)}, \label{eq:TotalHybridSlotFINALL}\\
H_{z_\text{II}}&=&\sin m\theta\sum\limits_{p=0}^{\infty}\frac{B_{p}}{Z_{0}}\sin\beta_{p}(z+\Delta)\frac{H_{m}^{(1)}(k_{tp}r)}{H_{m}^{(1)}(k_{tp}a)}, \label{eq:TotalHybridSlotFINALL2}\\
\bar{E}_{r_\text{II}}&=&\cos m\theta\sum\limits_{p=0}^{\infty}\frac{\sin\beta_{p}(z+\Delta)}{k^{2}_{tp}}\left[\frac{B_{p}}{Z_{0}} \frac{i\omega\mu m}{r}\frac{H_{m}^{(1)}(k_{tp}r)}{H_{m}^{(1)}(k_{tp}a)} - A_{p} \beta_{p} k_{tp}\frac{H_{m}^{'(1)}(k_{tp}r)}{H_{m}^{(1)}(k_{tp}a)}  \right], \label{eq:TotalHybridSlotFINALL3}\\
\bar{E}_{\theta_\text{II}}&=&\sin m\theta\sum\limits_{p=0}^{\infty}\frac{\sin\beta_{p}(z+\Delta)}{k^{2}_{tp}}\left[ -\frac{B_{p}}{Z_{0}} i\omega\mu k_{tp} \frac{H_{m}^{'(1)}(k_{tp}r)}{H_{m}^{(1)}(k_{tp}a)} + \frac{A_{p}\beta_{p} m}{r} \frac{H_{m}^{(1)}(k_{tp}r)}{H_{m}^{(1)}(k_{tp}a)}  \right], \label{eq:TotalHybridSlotFINALL4}\\
\bar{H}_{r_\text{II}}&=&\sin m\theta\sum\limits_{p=0}^{\infty}\frac{\cos\beta_{p}(z+\Delta)}{k^{2}_{tp}}\left[ \frac{B_{p}}{Z_{0}} \beta_{p} k_{tp} \frac{H_{m}^{'(1)}(k_{tp}r)}{H_{m}^{(1)}(k_{tp}a)} + \frac{A_{p}i\omega\epsilon m}{r} \frac{H_{m}^{(1)}(k_{tp}r)}{H_{m}^{(1)}(k_{tp}a)}  \right], \label{eq:TotalHybridSlotFINALL5}\\
\bar{H}_{\theta_\text{II}}&=&\cos m\theta\sum\limits_{p=0}^{\infty}\frac{\cos\beta_{p}(z+\Delta)}{k^{2}_{tp}}\left[  \frac{B_{p}}{Z_{0}}\frac{\beta_{p} m}{r}\frac{H_{m}^{(1)}(k_{tp}r)}{H_{m}^{(1)}(k_{tp}a)} + A_{p} i\omega\epsilon k_{tp} \frac{H_{m}^{'(1)}(k_{tp}r)}{H_{m}^{(1)}(k_{tp}a)}  \right], \label{eq:TotalHybridSlotFINALL6}
\end{eqnarray}
\end{widetext}

Using matrix conditioning and reduction steps similar to those used in \cite{Zotter}, we now take the following definitions to simplify our equations,
\begin{eqnarray} 
V_{n}&=&\frac{k_{0}}{k_{tn}}\frac{J'_{m}(k_{tn}a)}{J_{m}(k_{tn}a)}, \label{eq:DiagonalMatriced}\\
W_{n}&=&\frac{\beta_{n} m}{a k^{2}_{tn}}, \label{eq:DiagonalMatriced2}\\
X_{p}&=&\frac{k_{0}}{k_{tp}}\frac{H^{'(1)}_{m}(k_{tp}a)}{H^{(1)}_{m}(k_{tp}a)}, \label{eq:DiagonalMatriced3}\\
Y_{p}&=&\frac{\beta_{p} m}{a k^{2}_{tp}}, \label{eq:DiagonalMatriced4}\\
\hat{X}_{p}&=&\frac{(1+\delta_{p0})k_{0}}{k_{tp}}\frac{H^{'(1)}_{m}(k_{tp}a)}{H^{(1)}_{m}(k_{tp}a)}, \label{eq:DiagonalMatriced5}
\end{eqnarray}
where all the matrices in (\ref{eq:DiagonalMatriced})--(\ref{eq:DiagonalMatriced5}) are diagonal (either in $p$ or in $n$) and, therefore, easy to manipulate. The modal equations can now be written concisely as
\begin{eqnarray} 
W_{n}C_{n}+V_{n}D_{n}&=&\sum\limits_{p=0}^{\infty} \bar{M}_{np} (Y_{p}\bar{A}_{p}-X_{p}\bar{B}_{p}),\\
C_{n}&=&\sum\limits_{p=0}^{\infty}\bar{N}_{np}\bar{A}_{p},\\
\hat{X}_{p}\bar{A}_{p}-Y_{p}\bar{B}_{p}&=&\frac{b}{\Delta}\sum\limits_{n=-\infty}^{\infty}\bar{N}^{T}_{pn}(V_{n}C_{n}+W_{n}D_{n}),\\
-\bar{B}_{p}&=&\frac{b}{\Delta}\sum\limits_{n=-\infty}^{\infty}\bar{M}^{T}_{pn}D_{n},
\end{eqnarray}

In full matrix notation we can cast this into
\begin{eqnarray} 
WC+VD&=&\bar{M}Y\bar{A}-\bar{M}X\bar{B},\\
C&=&\bar{N}\bar{A},\\
\hat{X}\bar{A}-Y\bar{B}&=&\frac{b}{\Delta}(\bar{N}^{T}VC+\bar{N}^{T}WD),\\
-\bar{B}&=&\frac{b}{\Delta}\bar{M}^{T}D,
\end{eqnarray}
where $\bar{A}, \bar{B}, C, D$ are vector matrices, $\bar{N},\bar{M}$ are full real matrices and the remaining matrices are diagonal. We can further write this system of equations in terms of ``matrices of matrices" as follows
\begin{eqnarray} 
\begin{bmatrix} 1 & 0 \\ W & V \end{bmatrix} \begin{bmatrix} C \\ D \end{bmatrix}&=\begin{bmatrix} \bar{N} & 0 \\ 0 & \bar{M} \end{bmatrix}\begin{bmatrix} 1 & 0 \\ Y & -X \end{bmatrix}\begin{bmatrix} \bar{A} \\ \bar{B} \end{bmatrix}, \\
\begin{bmatrix} \hat{X} & -Y \\ 0 & -1 \end{bmatrix} \begin{bmatrix} \bar{A} \\ \bar{B} \end{bmatrix}&=\frac{b}{\Delta}\begin{bmatrix} \bar{N}^{T} & 0 \\ 0 & \bar{M}^{T} \end{bmatrix}\begin{bmatrix} V & W \\ 0 & 1 \end{bmatrix}\begin{bmatrix} C \\ D \end{bmatrix} ,
\end{eqnarray}
where $1$ and $0$ here represent the unity and zero matrices. We can solve these equations to eliminate the $\bar{A},\bar{B}$ pair. After manipulation, we obtain 
\begin{eqnarray} 
&&\begin{bmatrix} \bar{A} \\ \bar{B} \end{bmatrix}=\frac{b}{\Delta} \begin{bmatrix} \hat{X}^{-1} & -\hat{X}^{-1}Y \\ 0 & -1 \end{bmatrix} \begin{bmatrix} \bar{N}^{T} & 0 \\ 0 & \bar{M}^{T} \end{bmatrix}\begin{bmatrix} V & W \\ 0 & 1 \end{bmatrix}\begin{bmatrix} C \\ D \end{bmatrix}, \nonumber\\
&&\begin{bmatrix} 1 & 0 \\ W & V \end{bmatrix} \begin{bmatrix} C \\ D \end{bmatrix}=\frac{b}{\Delta}\begin{bmatrix} \bar{N} & 0 \\ 0 & \bar{M} \end{bmatrix}\begin{bmatrix} 1 & 0 \\ Y & -X \end{bmatrix}\nonumber \\
&& \ \ \ \times \begin{bmatrix} \hat{X}^{-1} & -\hat{X}^{-1}Y \\ 0 & -1 \end{bmatrix} \begin{bmatrix} \bar{N}^{T} & 0 \\ 0 & \bar{M}^{T} \end{bmatrix}\begin{bmatrix} V & W \\ 0 & 1 \end{bmatrix}\begin{bmatrix} C \\ D \end{bmatrix},
\end{eqnarray}

By appropriate grouping of the matrices in this homogeneous result we can finally write the system's characteristic equation explicitly in symmetric form as
\begin{eqnarray} 
&&0=\left(\begin{bmatrix} -V^{-1} & -WV^{-1} \\ -WV^{-1} & -W^{2}V^{-1}+V \end{bmatrix} -\right. \nonumber\\
&&\left. \frac{b}{\Delta} \begin{bmatrix} -\bar{N}\hat{X}^{-1}\bar{N}^{T} & -\bar{N}Y\hat{X}^{-1}\bar{M}^{T} \\ -\bar{M}Y\hat{X}^{-1}\bar{N}^{T} & -\bar{M}(Y^{2}\hat{X}^{-1}-X)\bar{M}^{T} \end{bmatrix} \right) \begin{bmatrix} C \\ D \end{bmatrix}, \label{eq:systemMatrix}
\end{eqnarray}
whose determinant must vanish to give the eigensolutions and the dispersion relation between $\omega$ and $\beta_{0}$, after we truncate the $p$ sum to some index $P_\text{max}$ and the $n$ sum to $\pm N_\text{max}$ in practice (remembering that the index $n$ runs in region 1, whilst the index $p$ runs in region 2). Specifically, we require that
\begin{eqnarray} 
&& 0=\text{Det}\left[\begin{bmatrix} -V^{-1} & -WV^{-1} \\ -WV^{-1} & -W^{2}V^{-1}+V \end{bmatrix}-\right. \nonumber\\
&&\left. \frac{b}{\Delta} \begin{bmatrix} -\bar{N}\hat{X}^{-1}\bar{N}^{T} & -\bar{N}Y\hat{X}^{-1}\bar{M}^{T} \\ -\bar{M}Y\hat{X}^{-1}\bar{N}^{T} & -\bar{M}(Y^{2}\hat{X}^{-1}-X)\mathbf{M}^{T} \end{bmatrix} \right],\label{determinant}
\end{eqnarray}
where 
\begin{eqnarray} 
\beta_{p}&=p\pi/(2\Delta), \ \ \  k_{tp}=\sqrt{k_{0}^{2}-\beta_{p}^{2}}, \label{definitons}\\
\beta_{n}&=\hat{\beta}_{0}+2\pi n/b,  \ \ \ \ k_{tn}=\sqrt{k_{0}^{2}-\beta_{n}^{2}},\label{definitons2}
\end{eqnarray}
\begin{eqnarray} \label{definitons_continued1}
V_{nn}&=\frac{k_{0}}{k_{tn}}\frac{J'_{m}(k_{tn}a)}{J_{m}(k_{tn}a)},  \ \ \ W_{nn}=\frac{\beta_{n} m}{a k^{2}_{tn}} \\
X_{pp}&=\frac{k_{0}}{k_{tp}}\frac{H^{'(1)}_{m}(k_{tp}a)}{H^{(1)}_{m}(k_{tp}a)},  \ \ \ Y_{pp}=\frac{\beta_{p} m}{a k^{2}_{tp}} \\
\hat{X}_{p}&=\frac{(1+\delta_{p0})k_{0}}{k_{tp}}\frac{H^{'(1)}_{m}(k_{tp}a)}{H^{(1)}_{m}(k_{tp}a)}
\end{eqnarray}
\begin{eqnarray} \label{definitons_continued2}
M_{np}&=\frac{2\beta_{p}}{b(\beta_{n}^{2}-\beta^{2}_{p})}\begin{cases}
\sin\beta_{n}\Delta, & \text{for even }p \\
-i\cos\beta_{n}\Delta, & \text{for odd }p \\
\end{cases} \\
N_{np}&=\frac{2\beta_{n}}{b(\beta_{n}^{2}-\beta^{2}_{p})}\begin{cases}
\sin\beta_{n}\Delta, & \text{for even }p \\
-i\cos\beta_{n}\Delta, & \text{for odd }p \\
\end{cases}
\end{eqnarray}

Solving (\ref{determinant}) numerically for a given frequency or wavenumber $k_{0}=\omega/c$ will produce the sought complex propagation constant $\beta_{0}$ of the iris-line, whose real part $\text{Re}[\beta_{0}]$ will represent the longitudinal phase constant and imaginary part $\text{Im}[\beta_{0}]$ the longitudinal attenuation constant.  Once we have found $\beta_{0}$, we can write the full fields by substituting into (\ref{TotalHybridLineFINALnormalized})--(\ref{eq:TotalHybridSlotFINALL6}).

\section{Numerical implementation and results}\label{results}

The mode-matching model obtained above is now ready for numerical implementation on a computer. Our investigation of the iris line is concerned not only in comparing the numerical results with the predictions of the model based on Vainstein's boundary condition, but also observing the influence of finite-screen thickness on the attractive characteristics (propagation loss, polarization and amplitude profile) of the dominant mode.  For the open iris-line structure in hand, we expect radiation loss and the paraxial propagation (similar to a plane wave, but with slightly bent, paraboloidal wavefronts \cite{Saleh}) to contribute, respectively, by giving a nonzero imaginary part of $\beta_{0}$ and a real part that is close but not exactly equal to $k_{0}$ [see equation (\ref{betaAppex}) in Appendix~\ref{Appx2} for a more formal description].

Assuming an incident THz wave with linear polarization, we are mainly interested in the dominant dipole mode on the iris line that can couple to and support such a wave. The basic aim of the numerical code is therefore to find the zeros of the determinant (\ref{determinant}) for a  field with $m=1$, which would give us $\beta_{0}=\text{Re}[\beta_{0}]+i\text{Im}[\beta_{0}]$. Given the periodicity of the dispersion curves in such a system, there will be an infinite number of possible solutions. The dominant mode will naturally be the one with the lowest attenuation constant (smallest $\text{Im}[\beta_{0}]$). Given the homogeneous equation (\ref{eq:systemMatrix}) of the $C_{n}, D_{n}$ field coefficients, our requirement for the determinant of this matrix to vanish is basically a requirement to find a value of $\beta_{0}$ that gives nontrivial field solutions. Substituting such a $\beta_{0}$ back in to the matrix will thus lower its rank and link the field coefficients $C_{n},D_{n}$ to each other, informing us about the ``shape" (or profile) of the dominant vector field distribution.  Solving this in practice using standard computer tools, such as Wolfram Mathematica (v.12.3.0), can be formulated as an eigenvalue problem. Indeed, the characteristic matrix (\ref{eq:systemMatrix}) and $[C, D]^{T}$ vector can be written in shorthand notation as $\mathsf{A}\mathsf{x} =0$, where $\mathsf{A}$ denotes the matrix  (\ref{eq:systemMatrix}), $\mathsf{x}$ the $[C, D]^{T}$ vector, and the superscript $T$ denotes taking the transpose. Then solving the homogeneous equation $\mathsf{A}\mathsf{x} =0$ is equivalent to solving the eigenequation $\mathsf{A}\mathsf{u} =\lambda_{u}\mathsf{u}$ to find the eigenvector $\mathsf{u}$ that corresponds to the zero eigenvalue (i.e.~if $\lambda_{u}=0$, then $\mathsf{x}\equiv\mathsf{u}$). Taking the $C_{n}$ and $D_{n}$ coefficients from this eigenvector and subtituting them, alongside $\beta_{0}$, into (\ref{TotalHybridLineFINALnormalized})--(\ref{eq:TotalHybridSlotFINALL6}) will then allow us to visualize the vector field polarization and profile, as will be shown in Subsection~\ref{Numericalresults}.

In the following subsections, we discuss the various aspects related to physical interpretation and practical implementation of the mode-matching method.

\hspace{0cm}	
\begin{figure}
\centering
\includegraphics[width=0.9\columnwidth]{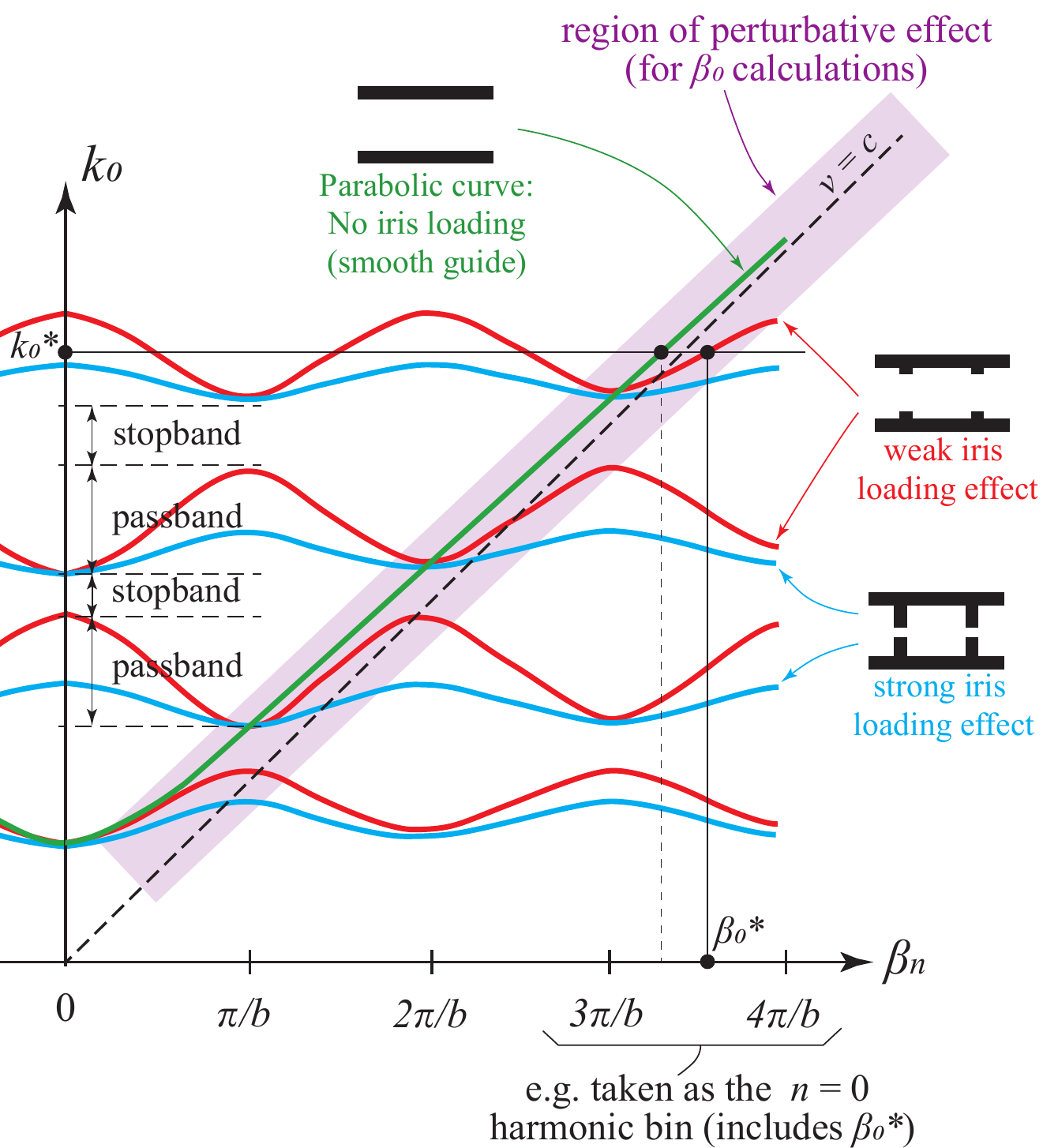}
\caption{A qualitative sketch demonstrating the effect of iris loading on an otherwise-smooth pipe as the depth of iris loading (added admittance) is changed. In this sketch $\beta_{n}$ is assumed to be real (no loss). For a smooth pipe, the continuous parabola (green) is observed, where phase velocity $v\geq c$ and $k_{0}\geq \beta_{0}$ for propagating modes, asymptotically leading to the limit of TEM propagation at the 45-deg line.  As we start loading the pipe with shallow (weak) irises, the propagated modes are allowed in certain passbands, attenuated in stopbands, and the dispersion curves (red) become periodic in $\beta_{n}$, with every side lobe period (index $n$) now corresponding to a Fourier harmonic bin, where $\beta_{n}=\beta_{0}+2\pi n/b$. This allows for slow and fast waves to propagate (depending on the used harmonic). If the irises are gradually removed, the passband curves (red) expand and coalesce back into the parabolic curves of the smooth pipe (green) \cite{slaterbook}. If the iris loading increases, the passbands (blue) are compressed, and in the limit they become flat spectral lines, with wider stopbands in-between. Since our structure is highly overmoded with a paraxial incidence, we effectively operate in the \emph{perturbation region} (purple shade) in the vicinity of the parabolic curve and make our choice for the $n=0$ harmonic band accordingly; see the simplified example of point $(\beta_{0}^{*},k_{0}^{*})$ on the figure.}   \label{fig:dispersioncurves}
\end{figure}

\subsection{Discussion on the truncation and clustering of field expansions}\label{clustering}

For numerical implementation, one clearly needs to truncate the infinite series in the $n$ and $p$ expansions seen in (\ref{TotalHybridLineFINALnormalized})--(\ref{eq:TotalHybridSlotFINALL6}). Ideally, the expansion in $n$ would be $\sum^{N_\text{max}}_{-N_\text{max}}$ and the expansion in $p$ would be $\sum^{P_\text{max}}_{0}$, where the value of $P_\text{max}$ is customarily chosen close to $2N_\text{max}$ (to facilitate matching both expansions across boundary, assuming thin screens) and both are high enough to allow for the solution to numerically settle at a given accuracy. Indeed, the value of $P_\text{max}$ can be estimated by intuitively considering how the field in region 2 is excited by the wave incident in region 1. Since we assume paraxial incidence ($\beta_{0}\approx k_{0}$)  from the THz source along the $z$ axis, and since the proposed structure is highly overmoded, with each period $b$ in the order of ${\sim}3000$ wavelengths, the longitudinal (guided) wavelength will roughly be equal to its value in free-space ${\sim}\lambda_{0}=\omega/c$. When this wave is matching tangentially across the boundary ($r=a$) towards region 2, whose width is $2\Delta$ (see Figure~\ref{fig:geometry}), we expect the excitation of the mode whose longitudinal standing-wave in region 2 has a half-wavelength value close to the value of ${\sim}\lambda_{0}/2$. This implies that the dominant mode number (call it $p=P_{0}$) in region 2 will be approximately the integer closest to $\frac{2\Delta}{\lambda_{0}/2}$. This, indeed, turns out to be the case, as can be observed through numerical computations. 

The modes in region 2, however, can be categorized, as in traditional smooth waveguides, to either propagating or evanescent modes. The former happens when $k_{0}>\beta_{p}$ (giving real $k_{tp}$, with phase velocity $v>c$ and $H^{(1)}_{m}(k_{tp}r)$ Hankel function radiation), while the latter corresponds to $k_{0}<\beta_{p}$ (giving purely imaginary $k_{tp}=i\gamma_{p}$, where $\gamma_{p}$ is a real number, with $v<c$ and $K_{m}(\gamma_{p}r)$ modified Bessel function radiation). The dispersion curve of such a waveguide takes the usual form of a parabola that is entirely above the 45-deg line (speed of light line), asymptotically approaching it from above in the high-frequency limit; see Figure~\ref{fig:dispersioncurves}.  Therefore, the integer $P_{0}$ closest to $\frac{2\Delta}{\lambda_{0}/2}$ must actually be chosen as the \emph{floor integer} (i.e.~the closest integer from below) to represent a \emph{propagating} mode (radiation) in region 2. Such a mode will typically be responsible for most of the attenuation seen in $\text{Im}[\beta_{0}]$.  If the ceiling integer was chosen instead, it would correspond to the largest evanescent mode.  Therefore, in our implementation and throughout the results below $P_{0}$ will refer to the floor integer of $\frac{2\Delta}{\lambda_{0}/2}=\frac{4\Delta}{\lambda_{0}}$. 

\begin{figure}
\centering
\includegraphics[width=\columnwidth]{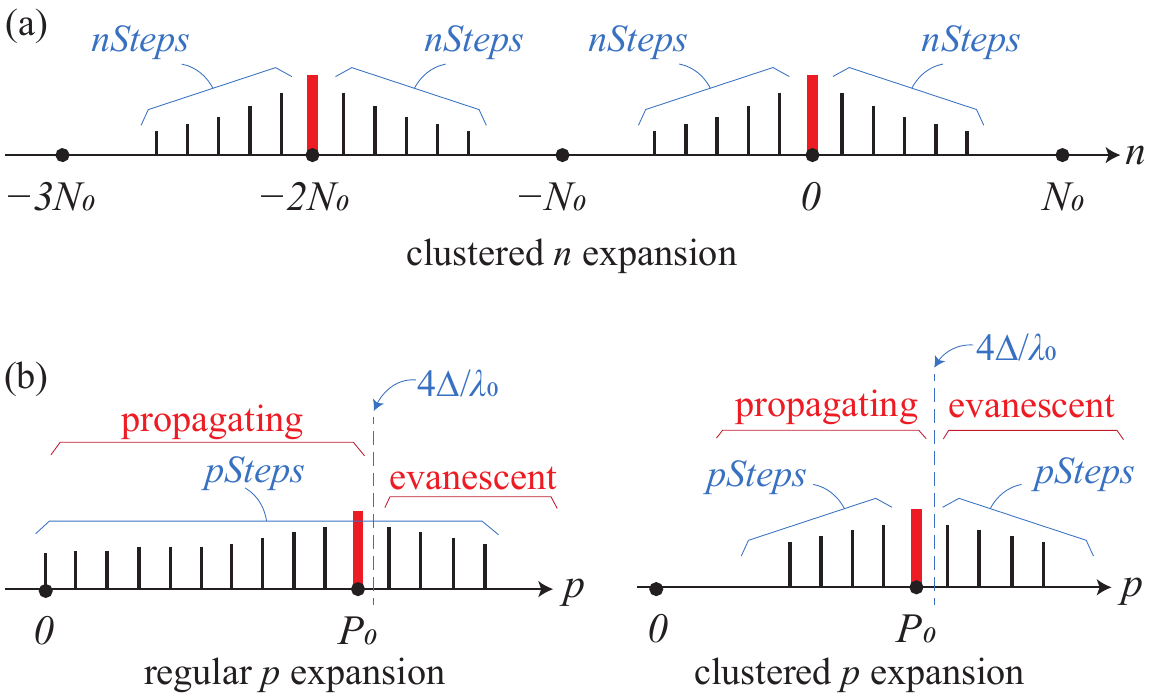}
\caption{(a) A clustered expansion in $n$, taking the paraxial dominant harmonic (in red) to be the basic harmonic ($\beta_{n}=\beta_{0}$) at $n=0$, which corresponds to approximately $N_{0}$ wavelengths per structure period, where $N_{0}$ is the nearest integer to $b/\lambda_{0}$. This means that the image of the dominant harmonic is at the index $n=-2N_{0}$ (in red). Note that the expansion is equivalent to $\sum^{nSteps}_{-nSteps}+\sum^{-2N_{0}+nSteps}_{-2N_{0}-nSteps}$, where we choose $nSteps$. (b) A regular versus clustered expansion in $p$, where the dominant mode in region 2 (excited by the paraxial incidence in region 1) happens at index $p=P_{0}$, where $P_{0}$ is the \textit{floor} integer nearest to $2\Delta/(\lambda_{0}/2)$ from below (i.e.~number of half-wavelengths in the width $2\Delta$). The regular expansion here is $\sum^{pSteps}_{p=0}$, and the clustered expansion is $\sum^{P_{0}+pSteps}_{p=P_{0}-pSteps}$, where we choose $pSteps$. }   \label{fig:clustering}
\end{figure}

The choice of $P_{0}$ also serves as a natural midpoint around which we may choose to ``cluster" (or localize) our $p$ expansion, so that, instead of summing over $\sum^{P_\text{max}}_{p=0}$, we sum over $\sum^{P_{0}+pSteps}_{P_{0}-pSteps}$, where we choose $pSteps$ during numerical implementation. A tradeoff clearly exists between higher accuracy of the results (higher $pSteps$) and the computational burden (time and memory) required to perform the calculation. Our choice to use, or not use, clustering will depend on how overmoded a structure is and how difficult it is to run a regular expansion (without clustering). Note that a regular expansion $\sum^{P_\text{max}}_{0}$ is equivalent to a clustered expansion $\sum^{P_{0}+pSteps}_{P_{0}-pSteps}$, if $pSteps$ happened to be equal to $P_{0}$ and $P_\text{max}$ happened to be equal to $2P_{0}$. Moreover, if a regular expansion up to $P_\text{max}$ stopped short of reaching $P_{0}$ (that is, if $P_\text{max}<P_{0}$), no correct solution will be found, because the dominant mode would not be included. This highlights the usefulness of clustered expansion when we can only expand in a limited number of terms (e.g.~due to computational limitations), since the clustered expansion will always include the dominant mode; see Figure~\ref{fig:clustering}. 

\hspace{0cm}	

A similar concept can be used to cluster the $n$ expansion around the dominant harmonic, which is the closest (rounded) integer (call it $n=N_{0}$) to the expected value $\frac{b}{\lambda_{0}}$. Unlike $P_{0}$, there is no need to seek the floor or ceiling integer values for $N_{0}$. Furthermore, we also need to include the Fourier image of the dominant harmonic. 

At this point, we are faced with a question as to where to look for the value of $\beta_{0}$ on the dispersion curves, when such curves are periodic in $\beta_{n}=\beta_{0}+2\pi n/b$? (see Figure~\ref{fig:dispersioncurves}). The periodicity naturally implies that all harmonic bins (lobes) are equivalent and we can choose to designate the index $n=0$ to any of them. However, since our structure is a highly-overmoded structure with a paraxial wavefront that can be considered a small perturbation compared to the TEM limit on the dispersion diagram, the dispersion curves forming the passbands in Figure~\ref{fig:dispersioncurves} will be in the vicinity of the parabolic curve of a smooth waveguide (small iris-loading effect). Indeed, in the limit of no iris loading (no diffraction losses), these passbands coalesce back to give the parabolic curve \cite{slaterbook}. For a perturbative iris effect, the dominant harmonic is expected to be near the corresponding $(k_{0},\beta_{0})$ point on the parabolic curve in Figure~\ref{fig:dispersioncurves}. Consequently, we choose $n=0$ to be the index for the dominant mode corresponding to $\beta_{n}=\beta_{0}\approx k_{0}$, which is equivalent to $N_{0}\approx b/\lambda_{0}$ harmonic bins away from zero on the $\beta_{n}$ axis. The Fourier image of the dominant mode will hence be at $n=-2N_{0}$, and a clustering of the form $\sum^{N_{0}+nSteps}_{N_{0}-nSteps}+\sum^{-2N_{0}+nSteps}_{-2N_{0}-nSteps}$ may be used for practical computations. 

The advantage of working around a dominant mode that is slightly perturbed (i.e.~clustering the field expansion) seems to, in some sense, offset the computational disadvantage of operating with such a highly-overmoded structure. Had the iris loading been too strong [e.g.~small Fresnel number $N_{f}=a^{2}/(b\lambda)$], more wave reflections between the irises would have resulted in several strong harmonic terms, which are not necessarily the original one near the parabolic curve (the periodic dispersion curves in Figure~\ref{fig:dispersioncurves} would be relatively flattened) and a preferred designation for the $n=0$ harmonic bin would no longer be clear. Good discussions on the interpretation of the $n$ harmonic bins and the limiting cases can be found in references \cite{slaterbook} and \cite{Hutter}.

The discussion above introduces the method of clustering in light of its intuitive nature in practice and the behavior infered from the $k_{0}-\beta$ dispersion curves. A more formal justification for using clustering in highly-overmoded paraxial lines is provided in Appendix~\ref{Appx2}, for completeness.

\subsection{Structure scales}\label{implementation}
Since the iris-line dimensions ($a, b$) under consideration are in the order of hundreds or thousands of wavelengths at THz frequencies, making the line highly overmoded, the computational cost experienced for mode-matching analysis and iterative optimization can be prohibitively high.  To help develop our understanding of overmoded iris-line structures and to help optimize our numerical algorithms, we utilize the clustering method where possible and we solve the problem over 3 scales. While keeping the aspect ratios fixed, each scale represents a similar structure with dimensions scaled 10-times relative to those of the next scale. Scale-1 dimensions are 100-times smaller than our intended structure (thus, 100 times less overmoded), Scale-2 is 10-times smaller than our intended structure, and Scale-3 is our intended structure; see Table~\ref{tab:scales}. All scales are analyzed at the same frequency of 3 THz. Clearly, solving the Scales 1 and 2 are less computationally prohibitive than Scale-3, where the structure period span thousands of wavelengths.  Working over scales will also allow us to spot different behavior trends as the scale moves from mildly-overmoded to highly-overmoded, as will be reported in Subsection~\ref{Numericalresults}. When validating Vainstein's model using mode-matching, it is important to keep an eye on the small parameter $M$ and the number $k_{0}b$ in each scale, since Vainstein's model assumes $M\ll 1$ and $k_{0}b\gg 1$, which imply a better validity in highly-overmoded structures. Thus, Scale-3 is expected to have stronger relevance to the predictions by Vainstein's model, compared to Scale-1 and Scale-2. 

\begin{table}[t]
\caption{\label{tab:scales}
The three different scales considered during the present analysis: mildly-overmoded (Scale-1), overmoded (Scale-2) and highly-overmoded (Scale-3). Scale-3 is the structure proposed for the LCLS THz transport problem. For each scale, $N_{0}$ indicates the structure period ($b$) measured in wavelengths, $N_{f}{=}a^{2}/(b\lambda_{0})$ is the Fresnel number and the values given for screen-thickness $\delta$ are example values to indicate the scaling. All shown dimensions are in mm. }
\begin{ruledtabular}
\begin{tabular}{ccccccc}
 Scale  & $a$ & $b$ & $\lambda_{0}$ & $\delta$ & $N_{0}({\approx}b/\lambda_{0})$ & $N_{f}$ \\
 \hline
1 & 0.55 & 3.33 & 0.1 & 0.01 & $\sim33\lambda_{0}$ & 0.9 \\
2 & 5.50 & 33.33 & 0.1 & 0.10 & $\sim333\lambda_{0}$ & 9.1 \\
3 & 55.00 & 333.33 & 0.1 & 1.00 & $\sim3333\lambda_{0}$ & 90.8 \\
\end{tabular}
\end{ruledtabular}
\end{table}

\begin{table}[t]
\caption{\label{tab:ALL}
Results from the mode-matching model for each scale, listing the corresponding Vainstein-based prediction (for $\delta=0$) for comparison and showing the influence of screen thickness. Note that the validity of the Vainstein predictions is better for $M\ll 1$ and $k_{0}b\gg 1$. The indicated values for $P_{0}$, $pSteps$ and $nSteps$ are given for the case with $\delta=0$. The values of $pSteps$/$nSteps$ are given in brackets under the Scale name in each heading. Note that $\Delta\text{Im}[\beta_{0}]/\text{Im}[\beta_{0}]$ here denotes the change in the imaginary part of $\beta_{0}$ relative to its value for zero-thickness screens. The values of $\beta_{0}, k_{0}$ are given in 1/m, while $a,b,\lambda_{0}$ and $\delta$ are given in mm.}
\begin{footnotesize}
\begin{ruledtabular}
\begin{tabular}{cc|c|ccccccc}
 \multicolumn{2}{c|}{\textbf{Scale-1}} &   Vainstein-based & \multicolumn{2}{c}{Mode-matching model}   \\
 \multicolumn{2}{c|}{(264/33)} &  $\text{Re}[\beta_{0}], \text{Im}[\beta_{0}]$ &  \multicolumn{1}{c|}{$\delta$} &  $\text{Re}[\beta_{0}], \text{Im}[\beta_{0}], \frac{\Delta\text{Im}[\beta_{0}]}{\text{Im}[\beta_{0}]}\%$   \\
 \hline
$a$     &  0.55  & 62732.2, 52.5  & \multicolumn{1}{c|}{0.00}  & 62725.5, 26.20, $-00.0\%$\\
$b$     & 3.33  &   &  \multicolumn{1}{c|}{0.01} &  62722.5, 31.64, $+20.8\%$\\
$\lambda_{0}$&  0.1     &   &  \multicolumn{1}{c|}{0.02} & 62725.23, 37.51, $+43.1\%$\\
$k_{0}$     &  62831.9     &   &  \multicolumn{1}{c|}{0.03} & 62738.97, 39.18, $+49.4\%$\\
$k_{0}b$    &   209.4     &   &  \multicolumn{1}{c|}{0.04} & 62728.80, 20.02 $-23.9\%$\\
$M$         &  0.21  &   &  \multicolumn{1}{c|}{0.05} & 62721.74, 24.68, $-5.9\%$\\
$N_{0}$     &    33   &   &  \multicolumn{1}{c|}{0.06} & 62719.89, 30.10, $+14.8\%$\\
$P_{0}$     &    66   &   &  \multicolumn{1}{c|}{0.07} & 62722.71, 36.00, $+37.3\%$\\
            &       &   &  \multicolumn{1}{c|}{0.08} & 62736.08, 38.35, $+46.3\%$\\
            &       &   &  \multicolumn{1}{c|}{0.09} & 62727.62, 20.05, $-23.5\%$\\
            &       &   &  \multicolumn{1}{c|}{0.10} & 62721.20, 24.41, $-06.9\%$\\
            &       &   &  \multicolumn{1}{c|}{0.15} & 62720.09, 22.96, $-12.4\%$\\
            &       &   &  \multicolumn{1}{c|}{0.20} & 62719.26, 22.90, $-12.6\%$\\
            &       &   &  \multicolumn{1}{c|}{0.25} & 62719.40, 22.00, $-16.1\%$\\
            &       &   &  \multicolumn{1}{c|}{0.30} & 62718.07, 21.10, $-19.5\%$\\
\hline
 \multicolumn{2}{c|}{\textbf{Scale-2}} &   Vainstein-based & \multicolumn{2}{c}{Mode-matching model}   \\
 \multicolumn{2}{c|}{(1332/333)} &  $\text{Re}[\beta_{0}], \text{Im}[\beta_{0}]$ &  \multicolumn{1}{c|}{$\delta$} &  $\text{Re}[\beta_{0}], \text{Im}[\beta_{0}], \frac{\Delta\text{Im}[\beta_{0}]}{\text{Im}[\beta_{0}]}\%$   \\
 \hline
$a$     &  5.50  & 62830.5, 0.166  & \multicolumn{1}{c|}{0.0}  & 62830.50, 0.1090, $-0.0\%$\\
$b$     & 33.33  &   &  \multicolumn{1}{c|}{0.1} &  62830.50, 0.1070, $-1.8\%$\\
$\lambda_{0}$&  0.1     &   &  \multicolumn{1}{c|}{0.2} & 62830.49, 0.1060, $-2.8\%$\\
$k_{0}$     &  62831.9     &   &  \multicolumn{1}{c|}{0.3} & 62830.48, 0.1050, $-3.7\%$\\
$k_{0}b$    &   2094.4     &   &  \multicolumn{1}{c|}{0.4} & 62830.48, 0.1042, $-4.6\%$\\
$M$         &  0.07  &   &  \multicolumn{1}{c|}{0.5} & 62830.48, 0.1038, $-4.8\%$\\
$N_{0}$     &    333   &   &  \multicolumn{1}{c|}{0.6} & 62830.49, 0.1035, $-5.0\%$\\
$P_{0}$     &    666   &   &  \multicolumn{1}{c|}{0.7} & 62830.48, 0.1030, $-5.5\%$\\
            &          &   &  \multicolumn{1}{c|}{0.8} & 62830.48, 0.1028, $-5.7\%$\\
            &          &   &  \multicolumn{1}{c|}{0.9} & 62830.48, 0.1025, $-6.0\%$\\
            &          &   &  \multicolumn{1}{c|}{1.0} & 62830.48, 0.1020, $-6.4\%$\\
            &          &   &  \multicolumn{1}{c|}{1.75} & 62830.47, 0.0966, $-11.4\%$\\
            &          &   &  \multicolumn{1}{c|}{2.5} & 62830.47, 0.0935, $-14.2\%$\\
            &          &   &  \multicolumn{1}{c|}{3.0} & 62830.47, 0.0936, $-14.1\%$\\
\hline
 \multicolumn{2}{c|}{\textbf{Scale-3}} &   Vainstein-based & \multicolumn{2}{c}{Mode-matching model}   \\
 \multicolumn{2}{c|}{(6666/3333)} &  $\text{Re}[\beta_{0}], \text{Im}[\beta_{0}]$ &  \multicolumn{1}{c|}{$\delta$} &  $\text{Re}[\beta_{0}], \text{Im}[\beta_{0}], \frac{\Delta\text{Im}[\beta_{0}]}{\text{Im}[\beta_{0}]}\%$   \\
 \hline
$a$     &  55.00  & 62831.8, 0.00052  & \multicolumn{1}{c|}{0}  & 62831.8, 0.000625, $-00.0\%$\\
$b$     & 333.33  &   &  \multicolumn{1}{c|}{1} &  62831.8, 0.000414, $-33.2\%$\\
$\lambda_{0}$&  0.1     &   &  \multicolumn{1}{c|}{2} & 62831.8, 0.000410, $-33.9\%$\\
$k_{0}$     &  62831.9     &   &  \multicolumn{1}{c|}{3} & 62831.8, 0.000400, $-35.4\%$\\
$k_{0}b$    &   20943.9     &   &  \multicolumn{1}{c|}{5} & 62831.8, 0.000397, $-36.0\%$\\
$M$         &  0.02  &   &  \multicolumn{1}{c|}{10} & 62831.8, 0.000385, $-37.9\%$\\
$N_{0}$     &    3333   &   &  \multicolumn{1}{c|}{25} & 62831.8, 0.000361, $-41.8\%$\\
$P_{0}$     &    6666   &   &  \multicolumn{1}{c|}{30} & 62831.8, 0.000362, $-41.6\%$\\
\end{tabular}
\end{ruledtabular}
\end{footnotesize}
\end{table}

\hspace{0cm}	
\begin{figure}
\centering
\includegraphics[width=\columnwidth]{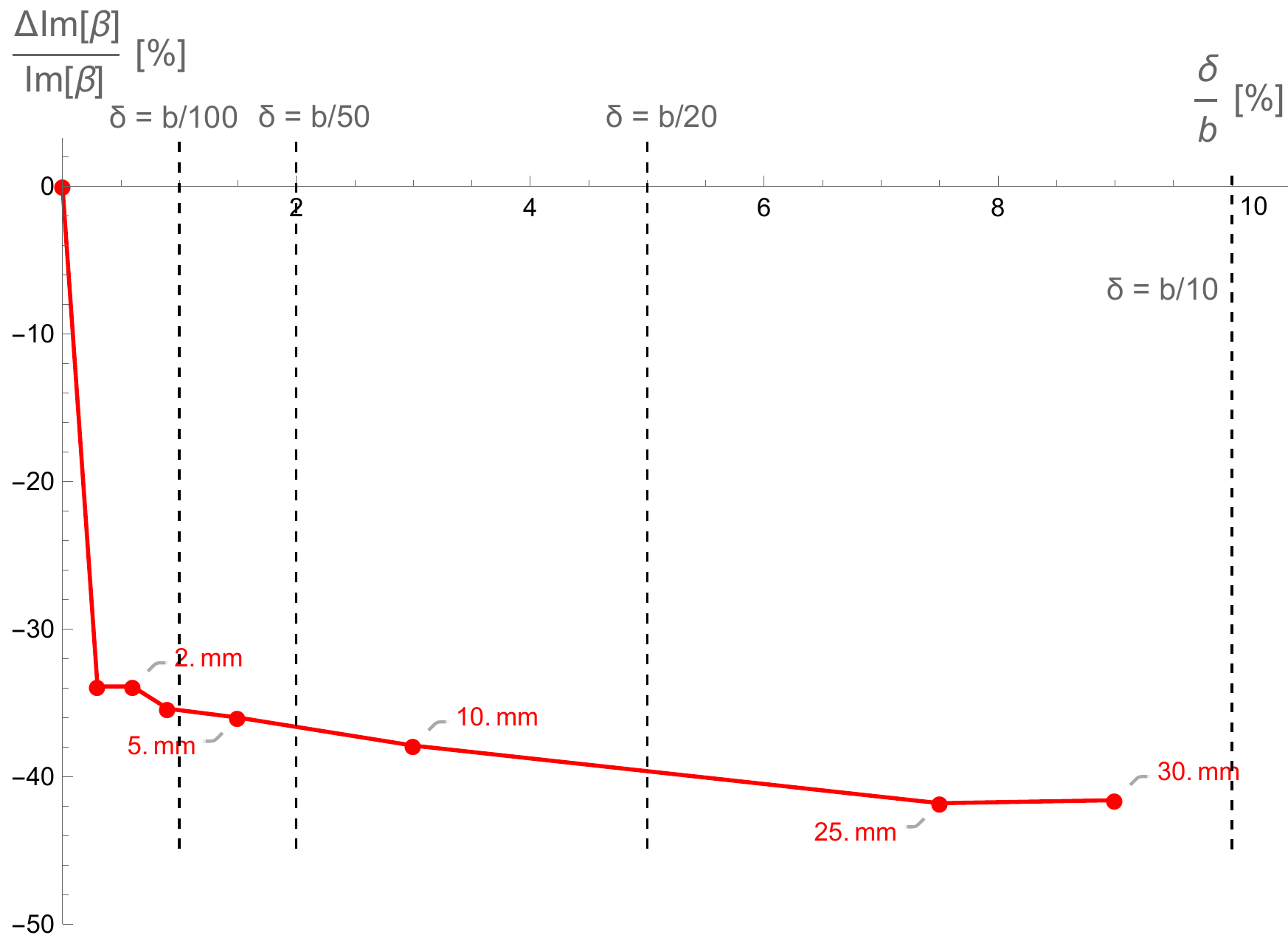}
\caption{Effect of varying screen thickness $\delta$ from zero to $\sim b/10$ (that is 0.0--30.0 mm) on $\text{Im}[\beta_{0}]$, for the Scale-3 structure with period $b=333.33$ mm, calculated at $pSteps/nSteps=6666/3333$. Selected points on the curve give explicit value of $\delta$ in mm. See data in Table~\ref{tab:ALL} for details.}   \label{fig:varyingthickness3}
\end{figure}

\subsection{Numerical results}\label{Numericalresults}
\begin{figure}
\centering
\includegraphics[width=\columnwidth]{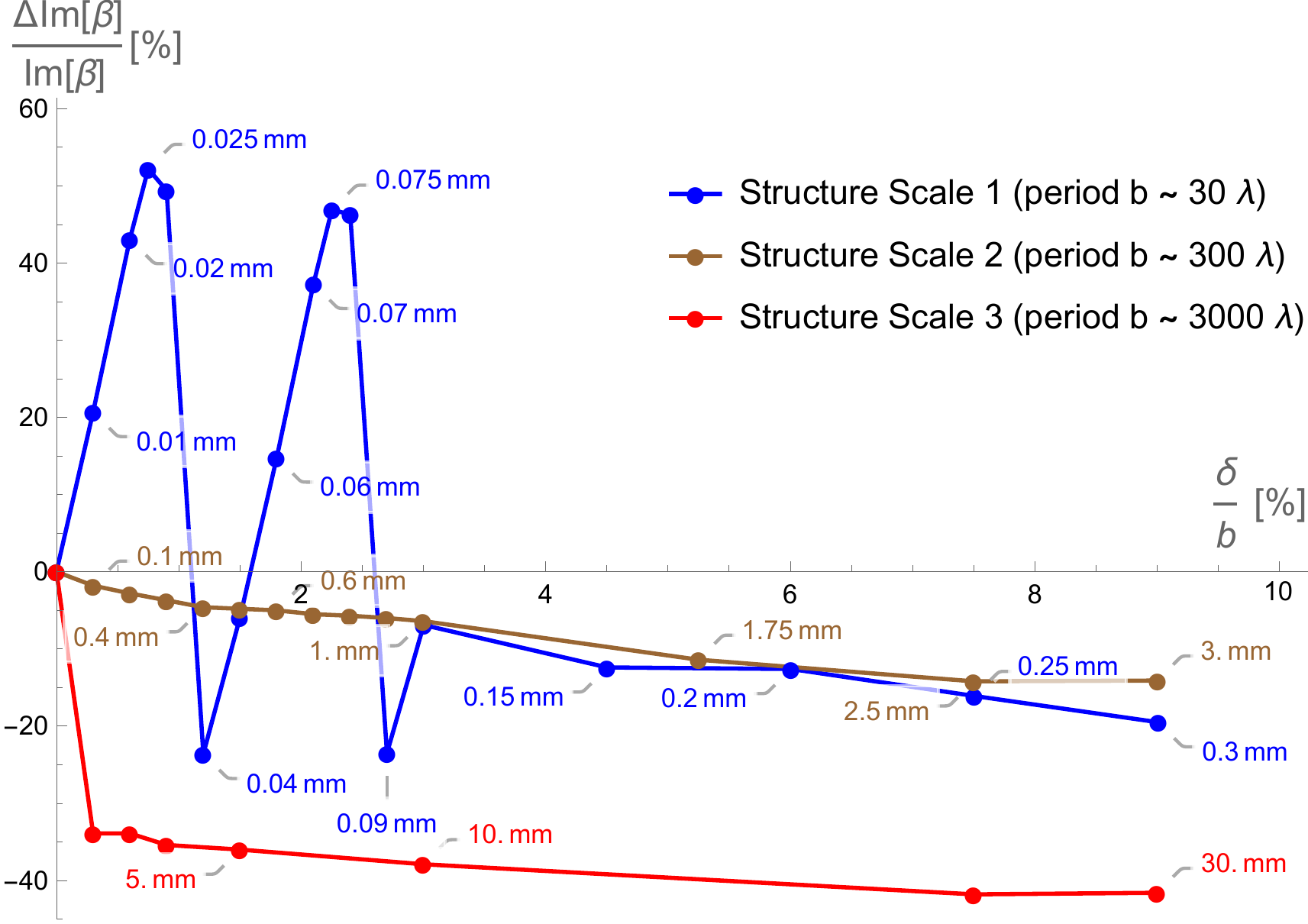}
\caption{Effect of varying screen thickness $\delta$ from zero to $\sim b/10$ on $\text{Im}[\beta_{0}]$, for the three scales discussed above, shown together on the same figure for comparison. Selected points on the curve give explicit value of $\delta$ in mm. See data in Table~\ref{tab:ALL} for  details.}   \label{fig:varyingthicknessALL}
\end{figure}

The results of mode-matching calculations are summarized in Table~\ref{tab:ALL} for the three scales of the iris line listed in Table~\ref{tab:scales}. In addition to comparing the mode-matching prediction with the Vainstein-based predictions for $\delta=0$ screens, the table shows the effect of increased screen thickness (varying $\delta$) on the attenuation constant ($\text{Im}[\beta_{0}]$), for the different scales. Following an iterative process, the solution was computed for a given $pSteps/nSteps$ choice, then re-computed for gradually-increased values of $pSteps$ and $nSteps$ in each iteration, until settlement is reached with less than ${\sim}0.5\%$ error. Table~\ref{tab:ALL} indicates the smallest $pSteps/nSteps$ values found for numerical settlement in each scale. Figures~\ref{fig:varyingthickness3} and \ref{fig:varyingthicknessALL} show how the attenuation loss constant is changing for the full-scale (Scale-3) structure as well as the two smaller scales. 

It is noted from Figure~\ref{fig:varyingthicknessALL} that the full-scale structure (Scale-3) will have $33.9\%$ reduction in the attenuation constant $\text{Im}[\beta_{0}]$ for a screen that is 2-mm thick. The change is less acute for Scale-2 structures around equivalent relative-thickness points ($\delta/b$). Scale-1 exhibits some oscillatory behavior for $\text{Im}[\beta_{0}]$ as a function of $\delta$ for relatively thin screens, before it settles in value for thicker screens. This may be attributed to the small Fresnel number of Scale-1, effectively taking the structure from the perturbative diffraction limit into stronger iris loading (admittance) where more harmonic terms experience stronger reflections and can interfere constructively and destructively, as implied by Figure~\ref{fig:dispersioncurves}. 
\hspace{0cm}	

\begin{figure}
\centering
\includegraphics[width=\columnwidth,angle=0]{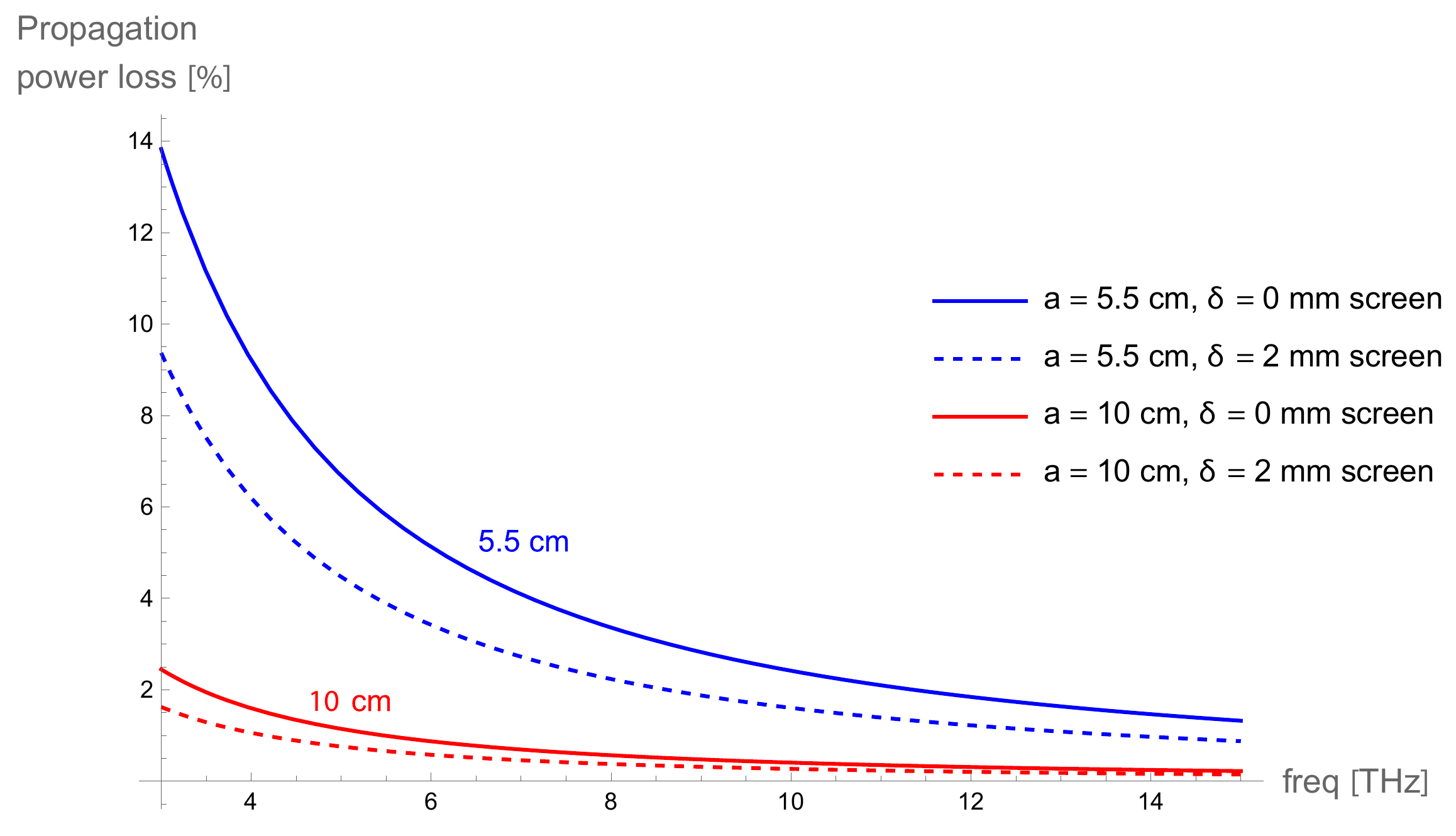}
\caption{Propagation loss is shown as a function of frequency for two iris radii ($a=5.5$ or $10$ cm) and two screen thicknesses (0 or 2 mm) and over a distance of 150 m to reach the Near Experimental Hall in LCLS. The iris-line period is $b=30$ cm and the effect of thickness is found from the mode-matching analysis and then used to refine Vainstein's model (assuming that it is the same effect for both cases, namely $\Delta\text{Im}[\beta_{0}]/\text{Im}[\beta]{\cong}-33.9\%$). The larger radius is recommended, if mechanically feasible and does not have a considerable transient at the line's entrance.}   \label{fig:LossCurveFinalAfter2mmWithTable}
\end{figure}

\begin{table}[t]
\caption{\label{tab:NEHandFEH}
Propagation power loss at 3 THz for the iris line with $b=30$ cm, two options of radius $a$, and over 150 m or 350 m transport distance. For LCLS, these are approximately the distances to the Near Experimental Halls (NEH) and Far Experimental Hall (FEH), respectively. The values are estimated using Vainstein's model after adjustment for screen thickness (assuming $\Delta\text{Im}[\beta_{0}]/\text{Im}[\beta]{\cong}-33.9\%$).}
\begin{ruledtabular}
\begin{tabular}{c|cccc}
 & \multicolumn{4}{c}{Propagation power loss at 3 THz}  \\
 \hline
Iris radius $a$ (mm) & \multicolumn{1}{c}{55} & \multicolumn{1}{c|}{55} & \multicolumn{1}{c}{100} & \multicolumn{1}{c}{100}\\
Screen thickness $\delta$ (mm) & \multicolumn{1}{c}{0} & \multicolumn{1}{c|}{2} & \multicolumn{1}{c}{0} & \multicolumn{1}{c}{2}\\
\hline
150-m path (for NEH) & \multicolumn{1}{c}{13.8\%} & \multicolumn{1}{c|}{9.4\%} & \multicolumn{1}{c}{2.4\%} & \multicolumn{1}{c}{1.6\%} \\
350-m path (for FEH) & \multicolumn{1}{c}{29.3\%} & \multicolumn{1}{c|}{20.5\%} & \multicolumn{1}{c}{5.6\%} & \multicolumn{1}{c}{3.7\%} \\
\end{tabular}
\end{ruledtabular}
\end{table}
Note that, for zero thickness screens, the predictions by the Vainstein-based model for $\text{Im}[\beta_{0}]$ deviate from those found numerically by about ${\sim}100\%$ for Scale-1 ($M=0.2$), ${\sim}52\%$ for Scale-2 ($M=0.07$) and ${\sim}17\%$ for Scale-3 ($M=0.02$). Clearly, the agreement is better in the limit of small $M$ (highly-diffractive/overmoded structure, with large Frensel number) where Vainstein's model is expected to increasingly hold. Within this validity limit, one may use the Vainstein-model for a quick first approximation of propagation loss and then add a correction to account for screen thickness effect, as predicted by the presented mode-matching results. For example, at 3 THz, the iris line originally proposed by reference \cite{DESY} for LCLS (with $a=5.5$ cm and $b=30$ cm) would give power loss as low as $13.8\%$ per 150 m (to reach the Near Experimental Hall at LCLS) using the Vainstein model and assuming infinitely thin screens (see Figure~\ref{fig:LossCurveFinalBefore}). When the screens are around 2-mm thick, for realistic implementation, we can adjust this value using the mode-matching predictions in Table~\ref{tab:ALL} as an approximate guide, reducing the final power loss estimate to $9.4\%$. If mechanically feasible, we can also use a larger iris radius of 10 cm (exploiting the $1/a^{3}$ dependence of loss factor on radius), to achieve an even lower loss of $2.4\%$ per 150 m for infinitely thin screens, which becomes $1.6\%$ for 2-mm thick screens (assuming similar $\delta$ effect as in the previous case); see Figure~\ref{fig:LossCurveFinalAfter2mmWithTable}.  

Using the same approach, Table~\ref{tab:NEHandFEH} summarizes propagation power loss values in the range of 3--15 THz for two lines: one with path length of 150~m to reach LCLS's Near Experimental Hall, while the other is for 350 m to reach the Far Experimental Hall.

\begin{figure*}
\centering
\includegraphics[width=\textwidth]{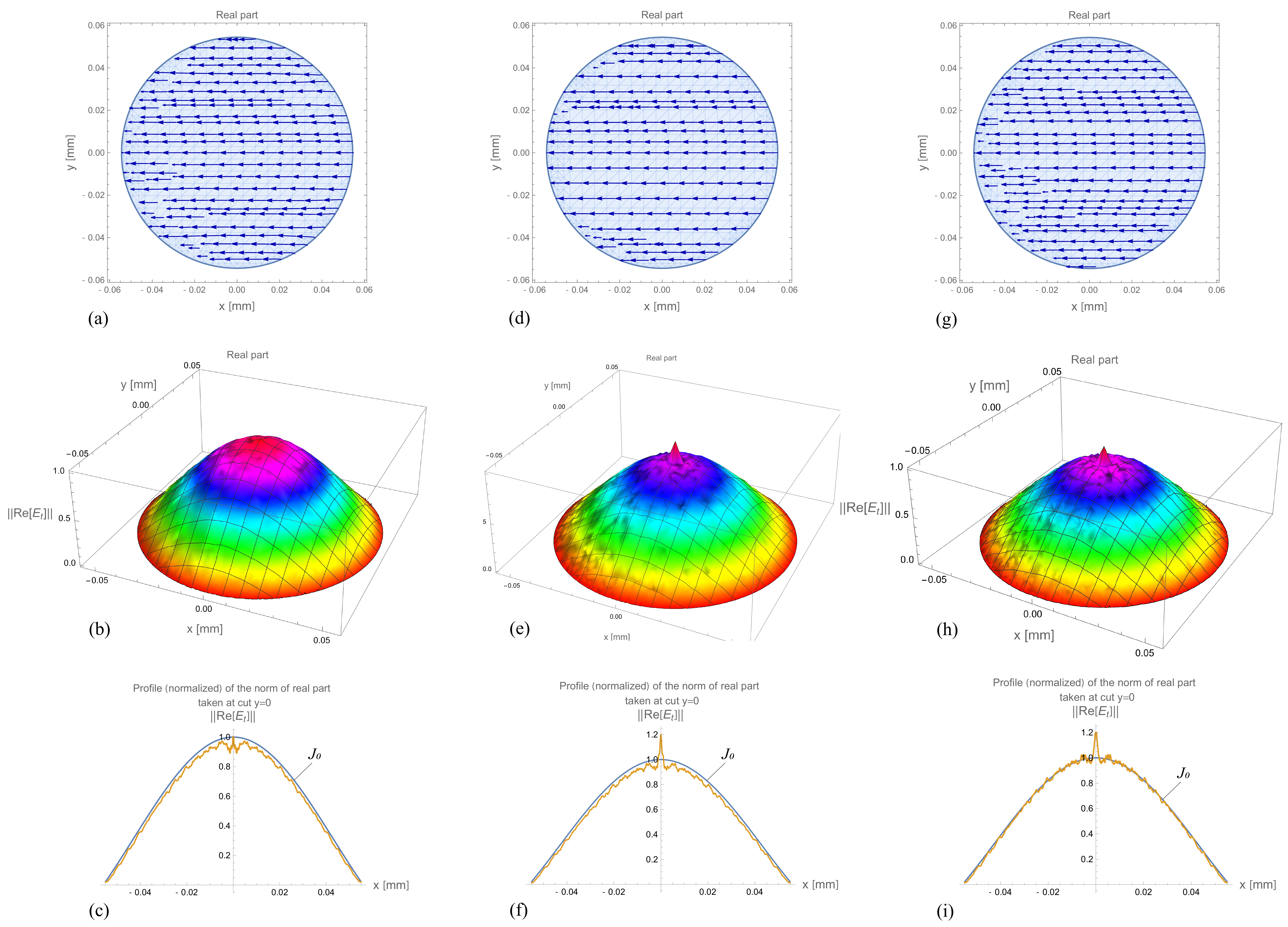}
\caption{Scale-3 field polarization and amplitude profile plots, calculated at $pSteps/nSteps=6666/3333$ and plotted at $nSteps=1111$: (a)--(c) for $\delta=0$ mm, (d)--(f) for $\delta=2$ mm, and (g)--(i) for $\delta=10$ mm. The polarization and amplitude profiles seem unaffected by the small increase in screen thickness. Subplots (c), (f) and (i) show $E$-field amplitude profile cuts at the $y=0$ plane across the iris, compared to the idealized shape of $J_{0}(2.4r/a)$ function. A small sharp spike is observed in the amplitude profile in (f) and (i), which may be a result of low numerical resolution (truncation).}   \label{fig:Scale3Summary}
\end{figure*}

The influence of screen thickness on field polarization and amplitude profile is shown in Figure~(\ref{fig:Scale3Summary}) for three indicative examples of screen thickness. The polarization (horizontal and invariant across the iris) and amplitude profile (closely resembling a $J_{0}(2.4r/a)$ function) seem to be unaffected by the thickness variation.

\hspace{0cm}	

\subsection{Ohmic losses}
Given the small skin depth of good conductors, such as copper or aluminum, at THz frequencies, we can use the well-known perturbation technique to compute the ohmic (conductor) losses at metallic screens, using our knowledge of the tangent magnetic field distribution near the screens when such screens are assumed to be perfectly conductive \cite{Pozar}. Specifically, the attenuation constant due to the conductive loss, $\alpha$, is calculated by the usual formula
\begin{equation}
    \alpha\cong\frac{P_{z}}{2P_{0}},
\end{equation}
where $P_{z}=\frac{1}{2} R_{s}\iint|H_\text{tan}|^{2}a d\theta dz$ is the power lost to conductive edges of the screen per unit length of the line, $P_{0}=\frac{1}{2}\text{Re}[\iint \mathbf{E}\times\mathbf{H}^{*}\cdot \hat{z}rdrd\theta]=\frac{1}{2}\text{Re}[\iint (E_{r}H_{\theta}^{*}-E_{\theta}H_{r}^{*})rdrd\theta]$ is the total power on the line, and $R_{s}=\sqrt{\omega \mu/(2\sigma)}$ is the surface resistance for a screen with conductivity $\sigma$. 
Finding the fields through the mode-matching equations (\ref{TotalHybridLineFINALnormalized})--(\ref{eq:TotalHybridSlotFINALL6}) then substituting in $\alpha$, we find that the ohmic losses on the iris line are negligible compared to the loss due to diffraction or radiation. For example, the attenuation constant $\alpha$ for a Scale-3 iris-line structure with 2-mm thick screens is found to be approximately $2.7\times 10^{-7}$ (1/m) for copper screens and $3.5\times 10^{-7}$ (1/m) for aluminum screens, which are more than 1000-times smaller than the attenuation constant $\text{Im}[\beta_{0}]{\cong}0.00041$ (1/m) due to diffraction (see Table~\ref{tab:ALL}).

\subsection{Limiting case test ($\Delta\rightarrow 0$)}\label{limtingcase}
As a compatibility check, we drove the current mode-matching model toward the limit of vanishing gap (approaching a smooth line) to compare the predicted eigenvalues with the  well-known values for a smooth circular waveguide. As expected, the hybrid mode decoupled into TE and TM modes in this limit, with the values of $\beta_{0}$ now approaching a purely real value (due to suppressed radiation by the closing gaps) and equal to the the analytical values of $\beta_\text{TE}=\sqrt{k^{2}_{0}-(V'_{mj}/a)^{2}}$ and $\beta_\text{TM}=\sqrt{k^{2}_{0}-(V_{mj}/a)^{2}}$, where $V'_{mj}$ is the $j$-th root of the $J'_{m}$ function, and  $V'_{mj}$ is the $j$-th root of the $J_{m}$ function, \cite{Pozar}. 

Specifically, using a Scale-1 structure with a gap as small as $2\Delta=10^{-5}$ mm, the mode-matching code gave $\beta_\text{TE}=62742.6114531+i 0.0023$ and $\beta_\text{TM}=62444.425897+i 0.0055$, compared to the analytic prediction (with at $2\Delta=0$) of $62742.611427+i 0$ and $62444.425854+i 0$, respectively. Notice the good agreement and that the imaginary parts given by the mode-matching are much smaller (negligible) compared to their values for the open gaps, which had $\text{Im}[\beta_{0}]\sim26$ (see Table~\ref{tab:ALL}).

\subsection{Transients, mode launching and other considerations}\label{limtingcase}
As was mentioned in Section~\ref{Intro}, the full implementation of this iris-line structure will require addressing further questions, in addition to those discussed above (propagation loss, polarization purity and amplitude profile). These questions pertain to the line dispersion, mode launching and matching from source to line, sensistivity to mechanical tolerances and fabrication misalignments, and the transient regime at the line's entrance (before the mode settles). Such questions are not the subject of this paper and will be examined in a subsequent publication by the authors. Here we briefly remark that mode launching is expected to have a strong influence on the length of the transient at the line's entrance. Indeed, if the incident wave is assumed to be an abstract plane wave, then reaching an equilibrium in the form of a dipole mode on the iris line is expected to be roughly around a distance ${\sim}\pi a^{2}/\lambda_{0}$ from the entrance (e.g.~see \cite{Palmer} and \cite{Lawson1}). However, if the field is launched using a specific mode from the source (e.g.~a Gaussian profile with linear polarization from a THz linear undulator), then the transient is expected to be much shorter.


\section{\label{Conclusions}Conclusions}
Vector field analysis using mode matching and clustered expansions has been presented for an overmoded iris-line structure for THz radiation transportation. The analysis includes the effect of finite screen thickness on the propagation properties of the dominant hybrid mode. As a specific application, the iris line was investigated as a method for THz radiation transport at LCLS, in the frequency range 3--15~THz. It is observed that having finite screen thicknesses, up to one tenth of the period of the iris-line ($\delta< b/10$), seems to reduce the attenuation constant $\text{Im}[\beta_{0}]$ down to approximately $-40\%$ for the highly-overmoded THz iris-line structure proposed for THz transport at LCLS. The horizontal polarization purity across the gap seem to be unaffected by the finite screen thickness. The field amplitude profile, which approximately follows a $J_{0}$ function profile (close to Gaussian), also seems to be unaffected by the finite screen thickness. The Vainstein-based model predictions, for zero-thickness screens, seems to hold increasingly in agreement with mode-matching numerical predictions for highly-diffractive/overmoded structures (limit of small $M$ parameter and large Fresnel numbers), but deviates considerably from numerical mode-matching predictions for mildly-overmoded structures or low Fresnel numbers. Finally, it is observed that the propagation losses on this type of iris line are dominated mainly by diffraction loss, rather than ohmic loss. 

\begin{acknowledgments}
The authors would like to thank Robert Warnock, Sami Tantawi, Emilio Nanni and Jeoff Neilson, of SLAC, Stanford University, for their encouragement and several interesting discussions in relation to this research. This work was supported
by the US Department of Energy (contract number DE-AC02-76SF00515).

\end{acknowledgments}

\appendix

\section{\label{Appx1}Derivation of power loss, field polarization and field intensity properties based on Vainstein's boundary condition}

In this appendix, we use perturbation theory to derive the vector field equations for the desirable dominant mode on the iris-line from Vainstein's boundary condition, highlighting its low propagation loss, linear polarization and amplitude profile. 

Vainstein's complex impedance condition on the virtual pipe (see Figure~\ref{fig:geometry1}) of the iris-line naturally implies that the general modes propagating in the line will be of the hybrid type (TE and TM modes couple through the impedance). The boundary condition of Vainstein at $r=a$ as formulated by Geloni, \emph{et al.} \cite{DESY}, can be written for the $E_{r}$ and $E_{\theta}$ field envelopes as
\begin{equation}
\left[ E + (1+i)\hat{\beta}_{0}aM \frac{\partial}{\partial r} E\right]_{r=a}={0}, \label{BCvectorial}
\end{equation}
where $\hat{\beta}_{0}=0.824$, as given in \cite{DESY,Vainstein1,Vainstein2}.

The hybrid-mode field equations are known for a regular pipe geometry and, if we drop the common factor $e^{i \beta z} e^{-i\omega t}$ for simplicity (i.e.~work in terms of envelopes only), can be written in terms of the transverse and longitudinal components as
\begin{eqnarray} 
E_{r}&=&\left[ \frac{i\beta}{k^{2}_{t}} E_{0} k_{t} J'_{n}(k_{t}r) {+} \frac{ik_{0} n}{k^{2}_{t}r} H_{0} J_{n}(k_{t}r)\right]\cos n\theta \label{fieldcomponents}\\
E_{\theta}&=&{-} \left[ \frac{i\beta n}{k^{2}_{t}r} E_{0} J_{n}(k_{t}r) {+} \frac{ik }{k^{2}_{t}} H_{0} k_{t} J'_{n}(k_{t}r)\right]\sin n\theta \label{fieldcomponents2}\\
E_{z}&=&E_{0}J_{n}(k_{t}r)\cos n\theta \label{fieldcomponents3}\\
H_{r}&=&\left[ \frac{i\beta }{k^{2}_{t}} H_{0} k_{t} J'_{n}(k_{t}r) {+} \frac{ik_{0} n}{k^{2}_{t}r} E_{0} J_{n}(k_{t}r)\right]\sin n\theta\label{fieldcomponents4}\\
H_{\theta}&=&\left[ \frac{i\beta n}{k^{2}_{t}r} H_{0} J_{n}(k_{t}r) {+} \frac{ik_{0} }{k^{2}_{t}} E_{0} k_{t} J'_{n}(k_{t}r)\right]\cos n\theta\label{fieldcomponents5}\\
H_{z}&=&H_{0}J_{n}(k_{t}r)\sin n\theta, \label{fieldcomponents6}
\end{eqnarray}
where $E_{0}$ and $H_{0}$ are constants.

In this overmoded structure, with relatively large Fresnel number, $N_{f}=\frac{a^{2}}{\lambda b}$, it is convenient to work in terms of the small parameter (in anticipation of perturbative analysis) that is inversely proportional to $N_{f}$. Such a parameter is $M=1/\sqrt{8\pi N_{f}}$, which was originally introduced in reference \cite{DESY}. Applying now the condition (\ref{BCvectorial}) to the $E_{r}, E_{\theta}$ fields in (\ref{fieldcomponents}), yields the following two relations
\begin{eqnarray}
0&=&J_{n}(k_{t}a)\left[ \frac{n}{a}H_{0}-\frac{n}{a^{2}}(1+i)\hat{\beta}_{0} a MH_{0} \right]+ \nonumber \\
&& J'_{n}(k_{t}a) \left[ k_{t}E_{0} +(1+i)\hat{\beta}_{0}aM\frac{n k_{t}}{a}H_{0} \right] + \nonumber \\
&& J''_{n}(k_{t}a)\left[ (1+i)\hat{\beta}_{0}aMk_{t}^{2} E_{0} \right] \label{eqInew}\\
0&=&J_{n}(k_{t}a)\left[ -\frac{n}{a}E_{0}+\frac{n}{a^{2}}(1+i)\hat{\beta}_{0} a ME_{0} \right]+\nonumber\\
&& J'_{n}(k_{t}a) \left[ -k_{t}H_{0} -(1+i)\hat{\beta}_{0}aM\frac{n k_{t}}{a}E_{0} \right] + \nonumber \\
&& J''_{n}(k_{t}a)\left[ -(1+i)\hat{\beta}_{0}aMk_{t}^{2} H_{0} \right] \label{eqIInew}
\end{eqnarray}

To nontrivially solve these two equations for the eigenvalue $k_{t}$ (the transverse wavenumber), let us multiply (\ref{eqInew}) by $E_{0}/H_{0}$ and then add/substrate (\ref{eqIInew}), respectively, to have
\begin{eqnarray}
&&0=\left( \frac{E^{2}_{0}}{H_{0}}{-}H_{0} \right) [J'_{n}(k_{t}a){+}(1{+}i)\hat{\beta}_{0}aMk_{t} J''_{n}(k_{t}a)] \label{star1}\\
&&0=\left( \frac{E_{0}}{H_{0}}{+}\frac{H_{0}}{E_{0}} \right) [J'_{n}(k_{t}a){+}(1{+}i)\hat{\beta}_{0}aMk_{t} J''_{n}(k_{t}a)] + \nonumber \\
&&  \frac{2n}{ak_{t}}J_{n}(k_{t}a){+}2(1{+}i)\hat{\beta}_{0}M[nJ'_{n}(k_{t}a){-}\frac{n}{ak_{t}}J_{n}(k_{t}a)] \label{star2}
\end{eqnarray}

Let's use perturbation theory in the small paramater $M$, and expand $k_{t}$ as 
\begin{equation}
k_{t}=k_{t_{0}}+c_{1}M+\mathcal{O}(M^{2}). \label{PTkt}
\end{equation}

For the unperturbed case when $M\rightarrow 0$, the conditions (\ref{star1}) and (\ref{star2}) reduce to 
\begin{eqnarray}
0&=&\left( \frac{E^{2}_{0}}{H_{0}}-H_{0} \right) J'_{n}(k_{t}a) \label{star1M=0}\\
0&=&\left( \frac{E_{0}}{H_{0}}+\frac{H_{0}}{E_{0}} \right) [J'_{n}(k_{t}a)+\frac{2n}{ak_{t}}J_{n}(k_{t}a)] \label{star2M=0}
\end{eqnarray}

From (\ref{star1M=0}), it is clear that we must either have $E_{0}=H_{0}$ or $E_{0}=-H_{0}$ as solutions. The former of these two is the desirable (balanced) hybrid mode that has a peak amplitude profile at the center of the iris and minimum at the walls (akin to the HE11 mode in traditional microwave corrugated waveguides \cite{Mahmoud,Clarricoats1,Clarricoats2}). The latter of the two, on the other hand, will result in a surface mode that ``sticks" to the walls, with a null at the center of the iris, and is undesirably lossy (akin to EH11 mode in traditional microwave corrugated waveguides \cite{Mahmoud,Clarricoats1,Clarricoats2}). Substituting in (\ref{star2M=0}) we immediately see how having $E_{0}=H_{0}$ in (\ref{star2M=0}) will result in $J_{n-1}(k_{t}a)=0$ with $k_{t_{0}}=V_{(n-1)j}/a$, whereas for $E_{0}=-H_{0}$ we have $J_{n+1}(k_{t}a)=0$ with $k_{t_{0}}=V_{(n+1)j}/a$, where $V_{nj}$ denotes the $j^{\text{th}}$ root (zero) of the Bessel function $J_{n}$. The dominant (lowest-order) modes in each of these cases are obtained when $n=1$ and give $J_{0}(2.4r/a)$ versus $J_{2}(5.1r/a)$ profiles, respectively. Note that in the case for $n=0$ it can be easily shown that the hybrid mode is degenerated to the TE or TM mode families. Let us now continue our treatment with the desirable balanced mode, with $E_{0}=H_{0}$, and for $\beta \approx k_{0}$ for paraxial high-frequency propagation above cutoff, and using perturbation theory to the first order in $M$ and with $k_{t_{0}}=V_{(n-1)j}/a$. 

Using the following Bessel identities \cite{NIST}
\begin{eqnarray}
J'_{n}(k_{t_{0}}a)+J_{n+1}(k_{t_{0}}a)&=&\frac{n}{k_{t_{0}}a}J_{n}(k_{t_{0}}a), \label{B1}\\
J'_{n}(k_{t_{0}}a)-J_{n-1}(k_{t_{0}}a)&=&\frac{-n}{k_{t_{0}}a}J_{n}(k_{t_{0}}a), \label{B2}
\end{eqnarray}
and exploiting the fact that our boundary condition has already forced $J_{n-1}(k_{t_{0}}a)=0$ in the unperturbed limit, we can deduce the following relations and perturbative expansions, up to $\mathcal{O}[M]$,
\begin{eqnarray}
J'_{n-1}(k_{t_{0}}a)&=&-J_{n}(k_{t_{0}}a), \label{B1final} \\
J'_{n}(k_{t_{0}}a)&=&-\frac{n}{k_{t_{0}}a}J_{n}(k_{t_{0}}a), \label{B2final}\\
J''_{n}(k_{t_{0}}a)&=&[J'_{n}(k_{t_{0}}a)]'\nonumber\\
&=&J_{n}(k_{t_{0}}a) \left[ \frac{n(1+n)}{a^{2}k^{2}_{t_{0}}} -1\right]. \label{B3final}    
\end{eqnarray}
\begin{eqnarray}
J_{n}(k_{t}a)&=&J_{n}(k_{t_{0}}a+c_{1}Ma)\nonumber\\
&=&J_{n}(k_{t_{0}}a)+c_{1}MaJ'_{n}(k_{t_{0}}a)+\cdots \nonumber \\
&\cong&J_{n}(k_{t_{0}}a)\left( 1-\frac{nc_{1}M}{k_{t_{0}}} \right) \label{expandJnnew} \\
J'_{n}(k_{t}a)&=&J'_{n}(k_{t_{0}}a+c_{1}Ma)\nonumber \\
&=&J'_{n}(k_{t_{0}}a)+c_{1}MaJ''_{n}(k_{t_{0}}a) \nonumber\\
&\cong&J_{n}(k_{t_{0}}a)\left[ c_{1}Ma\frac{n(n{+}1)}{a^{2}k^{2}_{t_{0}}}{-}ac_{1}M{-}\frac{n}{ak_{t_{0}}} \right] \label{expandJ'nnew} \\
J''_{n}(k_{t}a)&=&J''_{n}(k_{t_{0}}a+c_{1}Ma)\nonumber \\
&=&J''_{n}(k_{t_{0}}a)+c_{1}MaJ'''_{n}(k_{t_{0}}a)+\cdots \nonumber\\
&\cong&J_{n}(k_{t_{0}}a)\left[ \frac{n(n+1)}{a^{2}k^{2}_{t_{0}}}-1 \right]+c_{1}MaJ'''_{n}(k_{t_{0}}a)\nonumber\\
&&\text{\ \ } \label{expandJ''nnew} 
\end{eqnarray}

Substituting from (\ref{expandJnnew})--(\ref{expandJ''nnew}) into the original condition in (\ref{star2}), to first order in $M$, and using the shorthand notation $J_{n}(k_{t_{0}}a)\equiv J_{n}$ and $\left[ \frac{n(n+1)}{a^{2}k^{2}_{t_{0}}}-1 \right]\equiv \psi$, we find $c_{1}$ as follows
\begin{eqnarray}
&&0=[J'_{n}(k_{t}a)+(1+i)\hat{\beta}_{0}aMk_{t} J''_{n}(k_{t}a)] +\frac{n}{ak_{t}}J_{n}(k_{t}a) \nonumber\\
&& \ \ \  +(1+i)\hat{\beta}_{0}M[nJ'_{n}(k_{t}a)-\frac{n}{ak_{t}}J_{n}(k_{t}a)] \nonumber \\
&&=J_{n}\left[ -n - \frac{nc_{1}M}{k_{t_{0}}}+\psi \left( (1+i)\hat{\beta}_{0}a^{2}Mk^{2}_{t_{0}}+a^{2}k_{t_{0}}c_{1}M  \right)  \right] \nonumber \\
&&  + J_{n} n(1-\frac{nc_{1}M}{k_{t_{0}}}){+}(1{+}i)\hat{\beta}_{0}Mank^{2}_{t_{0}}\left( \frac{-n}{ak_{t_{0}}}J_{n}{+}c_{1}aM\psi J_{n} \right) \nonumber \\
&& \ \ \ \ \ \ -(1+i)\hat{\beta}_{0}Mk_{t_{0}}n \left[ J_{n}\left( 1-\frac{n c_{1}M}{k_{t_{0}}}\right) \right] \nonumber \\
&&=c_{1}M\left( a^{2} k_{t_{0}}\psi {-}n^{2}{-}n \right){+}(1{+}i)\hat{\beta}_{0}Mk_{t_{0}} \left( a^{2} k_{t_{0}}\psi {-}n^{2}{-}n \right) \nonumber \\
&&\Rightarrow c_{1}{=}{-}(1{+}i)\hat{\beta}_{0}k_{t_{0}}, \ \  k_{t}=\frac{V_{(n{-}1)j}[1{-}(1{+}i)\hat{\beta}_{0}M]}{a}, \label{c1foundnew}
\end{eqnarray}
where $n=1,2,3, \cdots$.  

The lowest-order (dominant) hybrid mode ($n=1, j=1$) will therefore have approximately the desired profile of a $J_{01}\left[ \frac{2.4 r}{a}[1-(1+i)\hat{\beta}_{0}M] \right]$ function. For small $M$, this is similar to the usual profile of the $J_{0}(2.4 r/a)$ function, but without allowing the ``skirt" of the $J_{0}$ function to go exactly to zero on the radius $r=a$ (as it would have in a smooth pipe). This allows the hybrid mode to exist at the boundary due to coupling (impedance) and gives the desired features of this mode of the iris line. The important result (\ref{c1foundnew}) agrees with that predicted by \cite{DESY} and is similar to the theory of the HE11 balanced hybrid-mode in traditional microwave corrugated guides (e.g.~see \cite{Mahmoud,Clarricoats1,Clarricoats2}). One should note, however, that the current model using Vainstein's boundary condition is mainly governed by diffraction phenomena, which dominates the propagation loss (\ref{VainsteinLoss}), rather than by ohmic (conductor) losses.

The propagation loss on the line can now be quantified using the imaginary part of $\beta$, as the attenuation constant. Using the fact that $\beta^{2}=k^{2}_{0}-k^{2}_{t}$ and the result in (\ref{c1foundnew}), we can now find the power loss ($L_{p}$) on the line by simply calculating $1-e^{-2\text{Im}[\beta]}$, which gives, after algebraic manipulation, the following
\begin{equation}
    L_{p}=\left[1-e^{-V^{2}_{mj}\hat{\beta}_{0}c^{3/2} b^{1/2} \omega^{-3/2}a^{-3}z} \right]\times 100\%, \label{VainsteinLoss}
\end{equation}
where $m\equiv n-1 = 0,1,2,\cdots$. Equation (\ref{VainsteinLoss}) indicates lower losses for higher frequencies (as a function of $\omega^{-1/2}$) and for larger radii (as a function of $a^{-3}$). Equation (\ref{VainsteinLoss__}) is a special case of (\ref{VainsteinLoss}), taken at $n=1,j=1$.

We can now proceed to examine the polarization of this desirable mode. For this balanced mode with $n\neq0$, we find that (\ref{fieldcomponents}) can be reduced to
\begin{eqnarray} 
E_{r}&=&\frac{i\beta }{k_{t}}E_{0} J_{n-1}(k_{t}r) \cos n\theta \label{Erfinal}\\
E_{\theta}&=&-\frac{i\beta }{k_{t}} E_{0} J_{n-1}(k_{t}r)\sin n\theta \label{Ephifinal}\\
H_{r}&=&\frac{i\beta }{k_{t}} E_{0} J_{n-1}(k_{t}r) \sin n\theta\\
H_{\theta}&=&\frac{i\beta }{k_{t}} E_{0} J_{n-1}(k_{t}r) \cos n\theta
\end{eqnarray}

To examine the $E$ field polarization, let us now convert equations (\ref{Erfinal}) and (\ref{Ephifinal}) to the Cartesian coordinates using the transformation $\hat{r}=\hat{x}\cos\theta+\hat{y}\sin\theta$ and $\hat{\theta}=-\hat{x}\sin\theta+\hat{y}\cos\theta$. After simplifying, this gives the following expression for the transverse field
\begin{eqnarray}
&&\mathbf{E}_{t}=\frac{ikE_{0}}{k_{t}}J_{n{-}1}(k_{t}r)\left[ \hat{r}\cos n\theta - \hat{\theta}\sin n\theta \right]\nonumber\\
&&=\frac{ikE_{0}}{k_{t}}J_{n{-}1}(k_{t}r)\left[ \hat{x}\cos(1{-}n)\theta {+} \hat{y}\sin(1{-}n)\theta  \right] \label{finalPol}
\end{eqnarray}
\hspace{0cm}	
\begin{figure}
\centering
\includegraphics[width=\columnwidth]{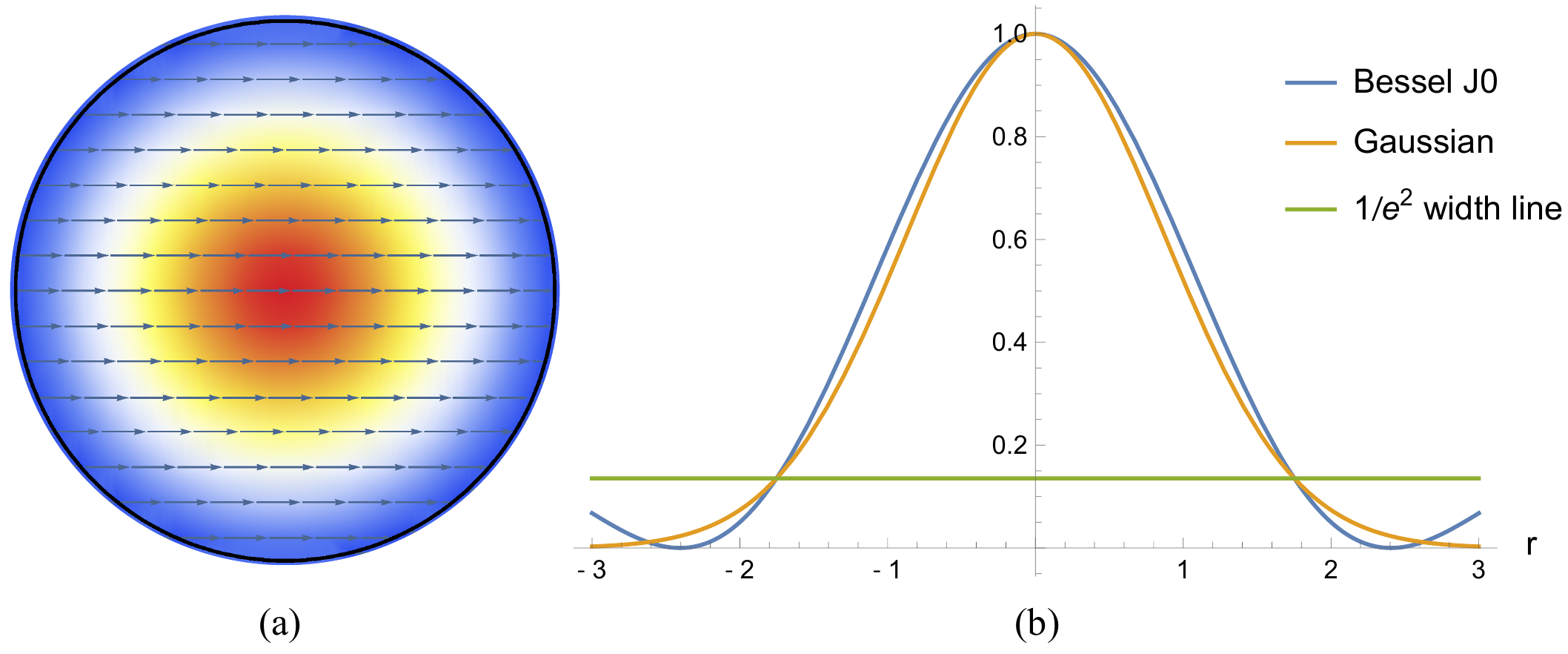}
\caption{(a) Transverse electric field stream lines and amplitude strength for the dominant hybrid mode ($n=1, j=1$) of the iris line at high-frequency operation (overmoded) above cutoff ($k_{0}\approx \beta $), based on equations (\ref{Erfinal}) and (\ref{Ephifinal}). This mode is promising for direct coupling with radiation from a THz linear undulator at LCLS. (b) A comparison between the intensity profiles of the Bessel function $J^{2}_{0}(r)$ and the Gaussian $e^{-r^{2}/\sigma^{2}}$, where $\sigma$ is the amplitude rms width.} \label{fig:Correctmode}
\end{figure}

\hspace{0cm}	
\begin{figure}
\centering
\includegraphics[width=\columnwidth]{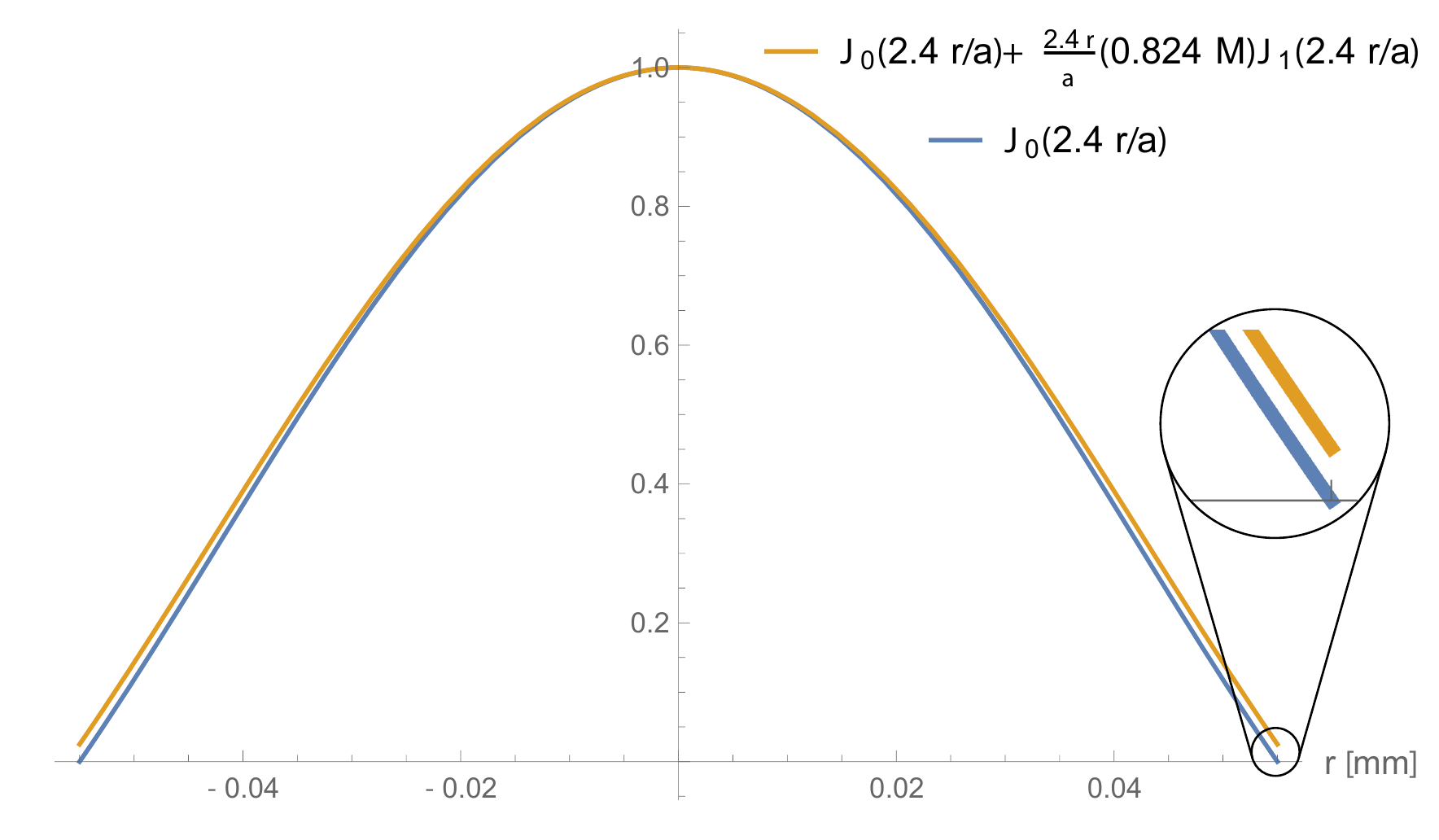}
\caption{A comparison between the real part of the field expansion (\ref{final11Taylor}) and the function $J_{0}(2.4r/a)$. The effect of having a small complex perturbation in the argument of the Bessel function will be to slightly lift the amplitude off zero near the edges, compared to the real function $J_{0}(2.4r/a)$.} \label{fig:secondTermReal}
\end{figure}

This result highlights the character of the polarization for the iris-line balanced modes. For the dominant mode of interest ($n=1, j=1$), (\ref{finalPol}) and (\ref{c1foundnew}) give
\begin{eqnarray}
\mathbf{E}_{x_{11}}&=&\hat{x}\frac{iak E_{0}J_{0}\left(\frac{2.4}{a}r[1-(1+i)\hat{\beta}_{0}M]\right)}{2.4 [1-(1+i)\hat{\beta}_{0}M]}, \\
                &\approx& \hat{x}\frac{iak E_{0}}{2.4}J_{0}\left(\frac{2.4}{a}r\right) \ \text{for } \ M\rightarrow 0, \label{final11} \\
\bar{E}_{y_{11}}&=&\hat{y} \ 0,
\end{eqnarray}
which means that the field is only polarized horizontally and it remains so across the entire aperture of the iris (fixed dipole polarization) while its amplitude profile is approximately $\approx J_{0}(2.4r/a)$. This gives an intensity profile that is similar to a Gaussian intensity (note: the profile of $J^{2}_{0}$ has about $97\%$ overlap with a perfect Gaussian intensity, when both are normalized to the same maximum and have same $1/e^{2}$ width below the maximum; see Figure~\ref{fig:Correctmode}b). Therefore, the dominant balanced mode of the iris-line is suitable for direct coupling with undulator radiation (see Figure~\ref{fig:Correctmode}a). As noted earlier, the complex argument of the Bessel function, with small complex perturbation proportional to $M$ will cause the edges of the function skirts to be slightly lifted off zero at $r=a$, allowing for the hybrid nature of the mode to be established. Indeed, this can be seen explicitly by writing out the real part of the final field expression. Since the Bessel's function $J_{n}$ is an \emph{entire} function (i.e.~analytic everywhere in the complex plane), the profile of $E_{x11}$ in (\ref{final11}) can be explicitly reduced further (with $M>0$) to its real and imaginary parts. For any fix $r$ value, one can Taylor expand in the small parameter $M$ to find that the field profile is proportional to
\begin{eqnarray}
E_{y_{11}}&=& \ 0,\\
E_{x_{11}}&\propto& J_{0}\left(\frac{2.4 r}{a}-\frac{2.4 r}{a}\hat{\beta}_{0}M(1+i)\right), \\
			&\propto& J_{0}\left(\frac{2.4 r}{a}\right)-\frac{2.4 r}{a}\hat{\beta}_{0}M(1+i)J^{'}_{0}\left(\frac{2.4 r}{a}\right)\nonumber\\
			&\propto& J_{0}\left(\frac{2.4 r}{a}\right)+\frac{2.4 r}{a}\hat{\beta}_{0}M J_{1}\left(\frac{2.4 r}{a}\right)\nonumber\\
			&&\ \ \ +i\frac{2.4 r}{a}\hat{\beta}_{0}M J_{1}\left(\frac{2.4 r}{a}\right) \label{final11TaylorTemp}
\end{eqnarray}

Thus, the real part of field profile approximately given as  
\begin{eqnarray}
E_{y_{11}}&=&\ 0,\\
E_{x_{11}}&\propto& J_{0}\left(\frac{2.4 r}{a}\right){+}\frac{2.4 r}{a}\hat{\beta}_{0}M J_{1}\left(\frac{2.4 r}{a}\right),\label{final11Taylor}
\end{eqnarray}
which is shown in Figure~\ref{fig:secondTermReal}. This shows the polarization and amplitude profile features of the dipole mode on the iris line. As one may expect, this mode is akin to dipole modes traditionally seen in similar periodic structures when driven by ultrarelativistic electron bunchs; e.g.~see \cite{Zotter,BaneWilson1}.

\hspace{0cm}	
\begin{figure}
\centering
\includegraphics[width=\columnwidth]{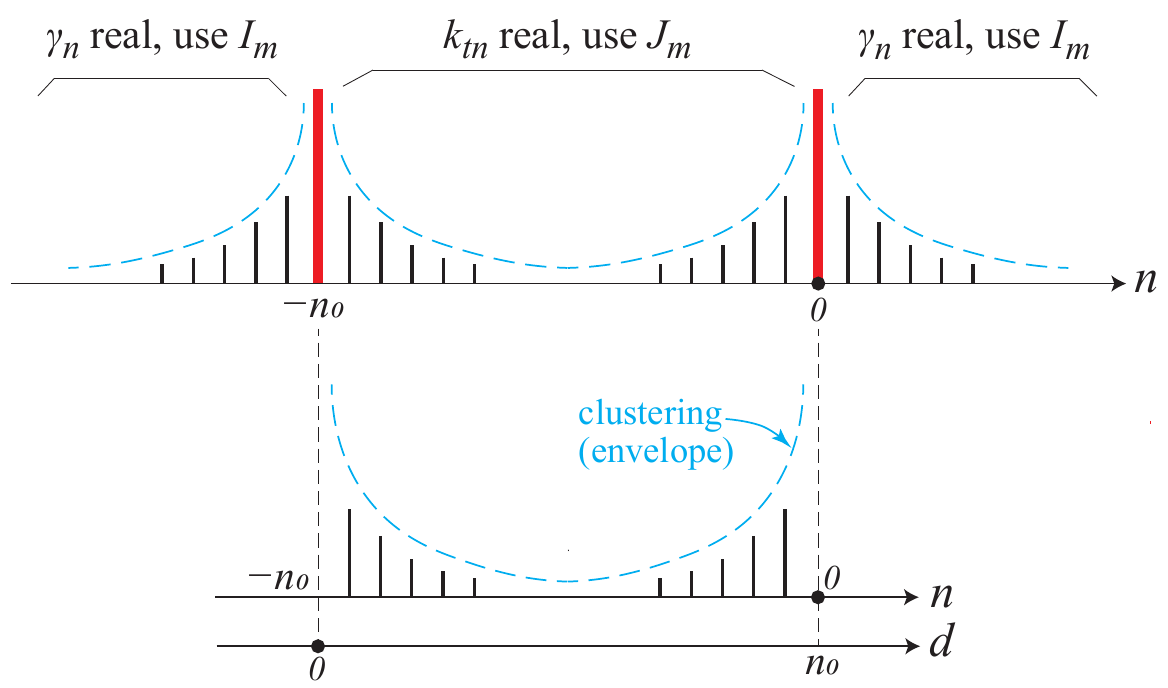}
\caption{An illustration of the harmonic distribution over the index $n$ under discussion (top axis), where the dashed line depicts the expected decay in harmonic strength as we move away from the band edges (in red) that represent the dominant harmonic and its image. Given the symmetry in the Bessel functions involved in the different bands, it is sufficient to focus our treatment on the middle band, between $-n_{0}$ and 0, where $k_{tn}$ is real. A convenient positive integer $d$ is introduced in the derivations, for the middle band, and is illustrated in the bottom axis. } \label{fig:clustersProof}
\end{figure}

\section{\label{Appx2}Formal justification for the use of clustering to approximate field expansions in an overmoded paraxial line}
In this appendix we use perturbation theory to derive a justification to the practice of clustering in field expansions, as used in Section~\ref{mode-matching} for a highly-overmoded iris line with paraxial incidence. Specifically, we wish to show that one can assume that the following two summations are interchangeable 
\begin{equation}
    \sum\limits_{-N_{\text{max}}}^{N_{\text{max}}} \longleftrightarrow \sum\limits_{-nSteps}^{nSteps}+\sum\limits_{-2N_{0}-nSteps}^{-2N_{0}+nSteps},
\end{equation}
where the first sum on the RHS is clustered around 0 (dominant harmonic, at $n{=}0$), while the second sum is clustered around its image (at $n{=}-n_{0}{=}-2N_{0}$), where $N_{0}$ is defined as the nearest integer to the number of wavelengths per structure period, $b/\lambda_{0}$; see Figure~\ref{fig:clustersProof}. Here $nSteps$ is chosen by us to provide a local (trunacted) span around the dominant harmonic and its image, such that the overall summation is shorter than the regular summation that would run continuously from $-N_{\text{max}}$ to $+N_{\text{max}}$, with the arbitrary index $N_{\text{max}}$ chosen high enough as to encompass all harmonics between and around the dominant harmonic and its image. The idea behind such a practice stems from an intuitive assumption that the dominant mode corresponds to an almost-TEM wave that is paraxially incident onto the overmoded iris line and slightly perturbed by the presence of the large irises, as discussed in Section~\ref{mode-matching}. Let's show that this assumption is formally justifiable. Without loss of generality, we specialize the discussion to the dipole mode ($m=1$) of the iris line.

We know that the propagation constant for the $n^{\text{th}}$ harmonic is  $\beta_{n}=\beta_{0}+2\pi n/b$, where $\beta_{0}$ approaches the wavenumber $k_{0}$ for paraxial propagation high above cutoff. In the limit of an ideal line (lossless with TEM wave), we can put $\beta_{n}=k_{0}+2\pi n/b$, which would lead to a transverse wavenumber $k_{tn}=\sqrt{k_{0}^{2}-\beta_{n}^{2}}=i\sqrt{(2\pi n/n)^{2}+4\pi nk_{0}/b}$. The dipole mode field equations at a given point along the line, say $z=0$, are given [from (\ref{TotalHybridLineFINALnormalized})--(\ref{TotalHybridLineFINALnormalized6})] by
\begin{eqnarray} 
E_{z_\text{I}}&=&\cos\theta \sum\limits_{n=-\infty}^{\infty}C_{n} \frac{J_{1}(k_{tn}r)}{J_{1}(k_{tn}a)}\label{TotalHybridLineFINALnormalized__}\\
H_{z_\text{I}}&=&\sin \theta \sum\limits_{n=-\infty}^{\infty}\frac{D_{n}}{Z_{0}} \frac{J_{1}(k_{tn}r)}{J_{1}(k_{tn}a)}\label{TotalHybridLineFINALnormalized2__}\\
E_{r_\text{I}}&=&i\cos \theta \sum\limits_{n=-\infty}^{\infty} \frac{D_{n}\omega\mu}{Z_{0}rk^{2}_{tn}}\frac{J_{1}(k_{tn}r)}{J_{1}(k_{tn}a)} {+} \frac{C_{n}\beta_{n}}{k_{tn}} \frac{J'_{1}(k_{tn}r)}{J_{1}(k_{tn}a)} \label{TotalHybridLineFINALnormalized3__}\\
E_{\theta_\text{I}}&=&i\sin \theta \sum\limits_{n=-\infty}^{\infty}  {-}\frac{D_{n}\omega\mu}{Z_{0}k_{tn}}\frac{J'_{1}(k_{tn}r)}{J_{1}(k_{tn}a)} {-}  \frac{C_{n}\beta_{n}}{r k^{2}_{tn}} \frac{J_{1}(k_{tn}r)}{J_{1}(k_{tn}a)} \label{TotalHybridLineFINALnormalized4__}\\
H_{r_\text{I}}&=&i\sin \theta \sum\limits_{n=-\infty}^{\infty}   \frac{D_{n}\beta_{n}}{Z_{0}k_{tn}}\frac{J'_{1}(k_{tn}r)}{J_{1}(k_{tn}a)} {+}  \frac{C_{n}\omega\epsilon}{r k^{2}_{tn}} \frac{J_{1}(k_{tn}r)}{J_{1}(k_{tn}a)} \label{TotalHybridLineFINALnormalized5__}\\
H_{\theta_\text{I}}&=&i\cos \theta \sum\limits_{n=-\infty}^{\infty} \frac{D_{n}\beta_{n}}{Z_{0}rk^{2}_{tn}}\frac{J_{1}(k_{tn}r)}{J_{1}(k_{tn}a)} {+}  \frac{C_{n}\omega\epsilon}{k_{tn}} \frac{J'_{1}(k_{tn}r)}{J_{1}(k_{tn}a)} \label{TotalHybridLineFINALnormalized6__}
\end{eqnarray}

It is clear from $k_{tn}=i\sqrt{(2\pi n/n)^{2}+4\pi nk_{0}/b}$ that certain $n$ values will lead to real $k_{tn}$ values, while other $n$ values will lead to purely imaginary $k_{tn}=i\gamma_{n}$, where $\gamma_{n}$ is real. For the former, the Bessel functions in (\ref{TotalHybridLineFINALnormalized__})--(\ref{TotalHybridLineFINALnormalized6__}) are Bessel functions of the first type [e.g.~$J_{1}(k_{tn}r)$], whereas for the latter case, they become, using the relation $J_{m}(i\gamma_{n}r)=i^{m}I_{m}(\gamma_{n}r)$, modified Bessel functions of the first type [e.g.~$I_{1}(\gamma_{n}r)$]. To work in terms of real $k_{tn}$ we will need negative $n$ values that satisfy $4\pi|n|k_{0}/b\geq 4|n|^{2}\pi^{2}/b^{2}$, which leads to the requirement that $-n_{0}\leq n \leq 0$, where $n_{0}=2b/\lambda_{0}=2N_{0}$. For the remaining values of $n$, $\gamma_{n}$ is real. See Figure~\ref{fig:clustersProof}.  Given the symmetry between the two types of formulations, we can work in either set of wavenumbers and Bessel functions ($k_{tn}$ and $J_{m}$, or $\gamma_{n}$ and $I_{m}$) and expect the other to have similar features. We can therefore confine the remainder of this discussion to the formulation using $k_{tn}$ and $J_{m}$ functions over the middle band shown in Figure~\ref{fig:clustersProof}. For the same iris-line dimensions used in Section~\ref{results}, we have $a=0.05$ m, $b=0.33$ m, $\lambda_{0}=10^{-4}$ m, which give $-6666=-n_{0}\leq n\leq 0$. We recognize the edges of this range as the terms we initially assumed to be the dominant harmonic and its image. Let's next show that the strength of all the harmonics between these two terms ($n=0,-n_{0}$) get smaller and smaller as we move away from edges.

Consider the total complex power associated with each harmonic travelling down the iris line, by calculating the integral (call it $I$) of the Poynting vector over the line's cross section,
\begin{eqnarray}
  I&=&\int\limits_{r=0}^{a}\int\limits_{\theta=0}^{2\pi}dr d\theta \ r (\mathbf{E}_{n}\times\mathbf{H}_{n}^{*})\cdot \hat{z}\nonumber\\
  &=&\int\limits_{0}^{a}\int\limits_{0}^{2\pi}dr d\theta \  r(E_{rn}H_{\theta n}^{*}-E_{\theta n}H_{r n}^{*}), \label{PoyntingEq}
\end{eqnarray}
where the asterisk denotes complex conjugation.  Before substituting the modal field expressions (\ref{TotalHybridLineFINALnormalized__})--(\ref{TotalHybridLineFINALnormalized6__}) into (\ref{PoyntingEq}) to find the respective powers carried by the harmonics, we should lift the idealistic assumption of lossless line, since the line will exhibit some loss. We introduce two perturbation parameters and write the complex propagation constant as
\begin{equation}
    \beta_{0}=k_{0}(1+\varepsilon_{1})+i \varepsilon_{2} k_{0},\label{betaAppex}
\end{equation}
where $\varepsilon_{1}{\ll}1$ and $\varepsilon_{2}{\ll}1$ represent, respectively, the small shift in real part of $\beta_{0}$ relative to $k_{0}$ and the small attenuation experienced by the waves on the line. Working to first order in the small parameters $\varepsilon_{1}, \varepsilon_{2}$, this leads, after algebraic manipulation, to
\begin{eqnarray}
\beta_{n}&=&\left[k_{0}(1+\varepsilon_{1})+\frac{2\pi n}{b}\right]+i[\varepsilon_{2} k_{0}] \label{beta_n}\\
|\beta_{n}|^{2}&\cong& k_{0}^{2}+\left(\frac{2\pi n}{b}\right)^{2}+2k_{0}\left[ \frac{2\pi n}{b}+\varepsilon_{1}(1+\frac{2\pi n}{b}) \right]\nonumber\\
&=&k_{0}^{2}+\phi^{2}_{n}, \\
\beta_{n}^{2}&\cong&k_{0}^{2}+\phi^{2}_{n}+2i\varepsilon_{2} k^{2}_{0}\left(1+ \frac{2\pi n}{b k_{0}} \right), \ \text{where}\\
\phi^{2}_{n}&=&\left(\frac{2\pi n}{b}\right)^{2}+2k_{0}\left[ \frac{2\pi n}{b}+\varepsilon_{1}\left(1+\frac{2\pi n}{b}\right) \right] \label{phi_n}
\end{eqnarray}

Within the present range of $n$, we define the convenient positive-integer measure $d$ as $d=n_{0}-|n|$, using it as our variable, and we substitute $n\rightarrow -|n|$ in (\ref{beta_n})--(\ref{phi_n}), to yield, after reduction to first order in $\varepsilon_{1}, \varepsilon_{2}$ and renaming $\phi_{n}$ as $\phi_{d}$,
\begin{eqnarray}
\phi_{d}&=&-\frac{4k_{0}^{2}}{n_{0}^{2}}d(n_{0}-d)-\frac{2 k_{0}^{2}}{n_{0}}\varepsilon_{1}(n_{0}-2d) \label{ee1}\\
\beta_{n}^{2}&\cong&k_{0}^{2}+\phi^{2}_{d}-i2\varepsilon_{2}k_{0}^{2}\left( \frac{n_{0}-2d}{n_{0}} \right) \\
k_{tn}^{2}&\cong&-\phi_{d}\left[ 1+\frac{i \varepsilon_{2}(n_{0}-2d)n_{0}}{2 d(n_{0}-d)+
\varepsilon_{1} n_{0}(n_{0}-2d)} \right]\\
k_{tn}&\cong&\psi_{1}+i\varepsilon_{2}\psi_{2}, \label{ktn}\ \ \text{where} \\
\psi_{1}&=&\sqrt{-\phi_{d}}\\
\psi_{2}&=&\frac{\sqrt{-\phi_{d}} (n_{0}-2d)n_{0}}{4 d(n_{0}-d)+
\varepsilon_{1} n_{0}(n_{0}-2d)}, 
\end{eqnarray}
noting that $\psi_{1},\psi_{2}$ are both real.

\hspace{0cm}	
\begin{figure}
\centering
\includegraphics[width=\columnwidth]{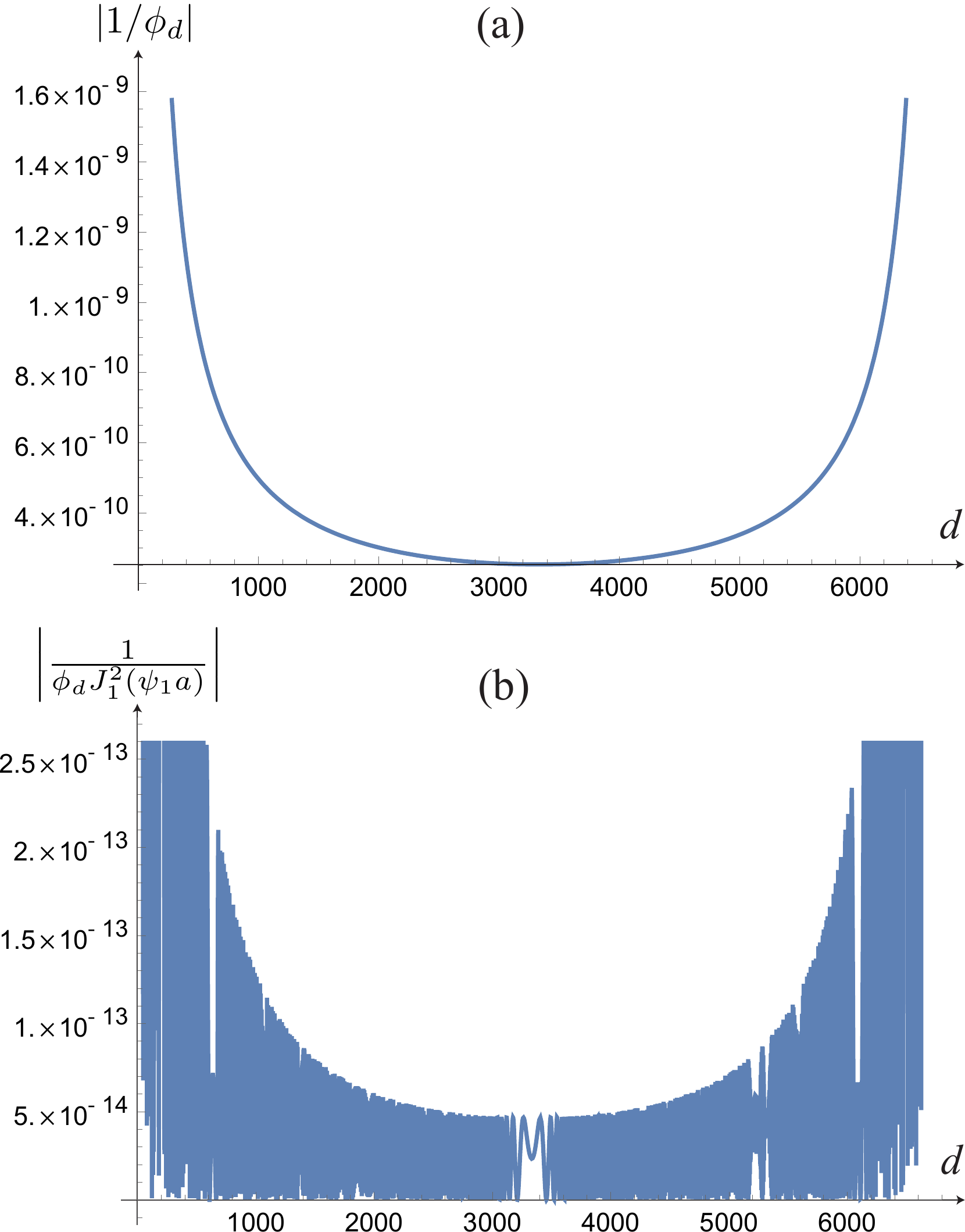}
\caption{(a) A plot of the convex modulation function, $|\phi_{d}^{-1}|$, responsible for the clustering effect, for the overmoded iris-line example with $a=0.055$ m, $b=0.33$ m, $\varepsilon_{1}=0.0001$ and $n_{0}$ of 6666. (b) An example of the effect of the modulation on the oscillatory term $1/J^{2}_{1}(\psi_{1}a)$, plotted using the formula (\ref{oscillatory}) for the same parameters.  The edges of the band (where $d$ tends to $0$ or $6666$), correspond to the dominant harmonic and its image; clustering is manifested in how the envelope of all harmonics decreases as we move away from band edges.} \label{fig:plotModulation}
\end{figure}

Substituting from the fields in (\ref{TotalHybridLineFINALnormalized__})--(\ref{TotalHybridLineFINALnormalized6__}) into the integral (\ref{PoyntingEq}), with the shorthand notation $J_{1}(k_{tn} r)\rightarrow J_{1}$, we now have (after algebraic reduction and integrating over $\theta$)
\begin{eqnarray}
I&=&\frac{\pi k_{0}}{2 Z_{0} |k_{tn}|^{2} |J_{1}(k_{tn}a)|^{2}}\left[\left( \beta^{*}_{n}|D_{n}|^{2}+\beta_{n}|C_{n}|^{2} \right) \int\limits^{a}_{0}dr \right.\nonumber\\
&&\left( r|J_{0}|^{2}+\frac{2}{r|k_{tn}|^{2}}|J_{1}|^{2}-2\text{Re}\left[ \frac{J_{1} J_{0}^{'*}}{k_{tn}}\right] \right)   \nonumber\\
&& + \left. \left( 2k_{0} \text{Re}[C_{n}D^{*}_{n}]+C_{n}D_{n}^{*}\frac{\phi^{'}_{n}}{k_{0}}\right) 2\text{Re}\left[  \int\limits^{a}_{0}dr\frac{J_{1} J_{1}^{'*}}{k_{tn}} \right] \right],\nonumber\\
&& \ \ \text{\ \ }\label{TotalIntegral}
\end{eqnarray}
where we identify four integrals to be performed, all of which involve Bessel functions with complex arguments. To find these integrals analytically, we convert the complex argument inside the Bessel functions into a real one by using (\ref{ee1})--(\ref{ktn}), Taylor expansions in the small imaginary parameter $\varepsilon_{2}$ (responsible for loss) and the standard Bessel identities for derivatives \cite{NIST}, to write
\begin{eqnarray}
J_{0}(k_{tn}r)&\cong&J_{0}(\psi_{1}r)-i\varepsilon_{2}\psi_{2}rJ_{1}(\psi_{1}r),\\
|J_{0}(k_{tn}r)|^{2}&\cong&J^{2}_{0}(\psi_{1}r), \\
J_{1}(k_{tn}r)&\cong&J_{1}(\psi_{1}r)\nonumber\\
&+&i\varepsilon_{2}\psi_{2} r\left[ J_{0}(\psi_{1}r) -\frac{1}{\psi_{1} r} J_{1}(\psi_{1}r)\right],\\
|J_{1}(k_{tn}r)|^{2}&\cong&J^{2}_{1}(\psi_{1}r), \\
J^{'}_{1}(k_{tn}r)&\cong&\left[ J_{0}(\psi_{1}r)-\frac{J_{1}(\psi_{1}r)}{\psi_{1}r}  \right]+i \varepsilon_{2}\frac{\psi_{2}}{\psi_{1}^{2}r}\left[ (1-\psi_{1}^{2}r^{2}\right.\nonumber \\
&&\  \left. +\psi_{1})J_{1}(\psi_{1}r)-\psi_{1}rJ_{0}(\psi_{1}r) \right],
\end{eqnarray}
which allow us to evaluate the four integrals. The result of integration can be summarized as
\begin{eqnarray}
&&\int\limits^{a}_{0}dr\left( r|J_{0}|^{2}+\frac{2}{r|k_{tn}|^{2}}|J_{1}|^{2}-2\text{Re}\left[ \frac{J_{1} J_{0}^{'*}}{k_{tn}}\right] \right)\nonumber\\
&& \ \ \ \ \ =\frac{a^{2}}{2} J^{2}_{0}(\psi_{1}a)+\left( \frac{a^{2}}{2}+\frac{1}{\phi_{d}}\right)J^{2}_{1}(\psi_{1}a),\\
&& 2\text{Re}\left[  \int\limits^{a}_{0}dr\frac{J_{1} J_{1}^{'*}}{k_{tn}} \right]=\frac{1}{\phi_{d}}\left[ 1-J^{2}_{1}(\psi_{1}a) \right].
\end{eqnarray}

We can now substitute back into (\ref{TotalIntegral}), noticing that we can approximate $|k_{tn}|^{2} |J_{1}(k_{tn}a)|^{2}\cong\-\phi_{d}J^{2}_{1}(\psi_{1}r)$, to finally yield the power in the $n^{\text{th}}$ harmonic as
\begin{eqnarray}
I&\cong& \frac{-\pi k_{0}}{2Z_{0}\phi_{d}}\left\{ \left( \beta^{*}_{n}|D_{n}|^{2}+\beta_{n}|C_{n}|^{2} \right) \left[ \frac{a^{2}}{2} \frac{J^{2}_{0}(\psi_{1}r)}{J^{2}_{1}(\psi_{1}a)} \right. \right.\nonumber\\
&& \left.   +\frac{a^{2}}{2}+\frac{1}{\phi_{d}}\right] +  2\text{Re}[C_{n}D^{*}_{n}]\frac{k_{0}}{\phi_{d}}\left[ \frac{1}{J^{2}_{1}(\psi_{1}a)} -1\right] \nonumber\\
&&\left. +C_{n}D^{*}_{n}\frac{1}{k_{0}}\left[ \frac{1}{J^{2}_{1}(\psi_{1}a)}-1 \right]  \right\} \label{FinalIntegralResult}
\end{eqnarray}

This result highlights how all the terms on the RHS will be modulated by the function $\phi_{d}^{-1}$, which is a convex function of $d$ that decays as we move towards the midpoint of the band. Thus, this function is responsible for local harmonic power decay as we move away from the dominant harmonic and its image (at the band edges where $d$ tends to $0$ and $n_{0}$); see Figure~(\ref{fig:clustersProof}).  Using (\ref{ee1}), we can plot this convex modulating function as shown in Figure~\ref{fig:plotModulation}a. Note that even though some of the terms in the RHS of (\ref{FinalIntegralResult}) contain the function $1/J^{2}_{1}(\psi_{1}a)$, which is highly oscillatory for an overmoded structure, their envelope will still be modulated by the function $\phi^{-1}_{d}$ to exhibit lower power strengths as their harmonic index $n$ (or $d$) moves away from dominant harmonic and its image. Indeed, this can be conveniently visualized by explicitly taking the asymptotical form of $J_{1}$ for large argument \cite{NIST}, $J_{1}(\psi_{1}a)\sim\sqrt{2/(\pi \psi_{1}a)}\sin(\psi_{1}a-\pi/4)$, then perturbatively expanding this form as well as the $\psi_{1}$ function, to first order in the small parameter $\varepsilon_{1}$, to give, after algebraic manipulation, 
\begin{eqnarray}
\psi_{1}&\cong& \frac{2k_{0}}{n_{0}}\sqrt{d(n_{0}-d)}\left[ 1+\varepsilon_{1}\frac{n_{0}(n_{0}-2d)}{4 d(n_{0}-d)} \right],\\
\frac{1}{J^{2}_{1}(\psi_{1}a)}&\cong&\frac{n_{0}\left[ 1-\varepsilon_{1}\frac{n_{0}(n_{0}-2d)}{4d(n_{0}-d)} \right]}{2\pi a k_{0}\sqrt{d(n_{0}-d)}}\left( 1-\sin\frac{4k_{0}a}{n_{0}}\sqrt{d(n_{0}-d)}\right)\nonumber\\
&& \ \ \ -\varepsilon_{1} \frac{n_{0}}{2\pi}\frac{n_{0}-2d}{d(n_{0}-d)}\cos\frac{4 k_{0} a}{n_{0}}\sqrt{d(n_{0}-d)}  \label{oscillatory}
\end{eqnarray}

Figure~\ref{fig:plotModulation}b uses (\ref{oscillatory}) to plot the modulated oscillatory term $\left[\phi_{d}J^{2}_{1}(\psi_{1}a)\right]^{-1}$, showing the clustering envelope over the oscillating terms.

\hfill


\bibliography{myrefs}

\end{document}